\documentclass[prx,10pt,twocolumn,superscriptaddress,floatfix, showpacs,nofootinbib]{revtex4-2}

\usepackage[utf8]{inputenc}  
\usepackage[T1]{fontenc}     
\usepackage[british]{babel}  
\usepackage[sc,osf]{mathpazo}
\usepackage[scaled=0.86]{berasans}  
\usepackage[colorlinks=true, allcolors=blue, urlcolor=blue]{hyperref}  
\usepackage{graphicx} 
\usepackage[babel]{microtype}  
\usepackage{amsmath,mathtools,amssymb,amsthm,bm,amsfonts,mathrsfs,bbm} 

\usepackage{xspace}  
\usepackage{pgfplots}
\usepackage{xcolor,colortbl}
\usepackage{array}
\usepackage{bigstrut}
\usepackage{mathrsfs}
\usepackage{svg}
\usepackage{colortbl}  
\usepackage{booktabs}

\usepackage{caption}
\usepackage{subcaption}
\usepackage{ragged2e}
\DeclareCaptionJustification{justified}{\justifying}
\usepackage{tabularx}
\newcolumntype{C}{>{\centering\arraybackslash}X}

\usepackage{lipsum}

\usepackage{verbatim}

\usepackage{comment}

\newcommand{\ar}{\color{black}}

\newcommand{\R}{\mathcal{R}}
\newcommand{\N}{\mathbb{N}}

\newcommand{\E}{\mathcal{E}}

\newcommand{\tr}{\text{tr}}
\newcommand{\ket}[1]{| #1 \rangle}

\newcommand{\be}{\begin{equation}}
\newcommand{\ee}{\end{equation}}
\newcommand{\bea}{\begin{eqnarray}}
\newcommand{\eea}{\end{eqnarray}}
\newcommand{\bes}{\begin{equation*}}
\newcommand{\ees}{\end{equation*}}
\newcommand{\beas}{\begin{eqnarray*}}
	\newcommand{\eeas}{\end{eqnarray*}}

\def\N{\mathcal{N}}

\def\tr{\mathrm{tr}}

\newtheorem{thm}{Theorem}
\newtheorem{cor}{Corollary}
\newtheorem{lem}{Lemma}
\newtheorem{prop}{Proposition}
\newtheorem{defn}{Definition}

\begin{document}
  
\title{ Nonlocal and quantum advantages in network coding for multiple access channels }

\author{Jiyoung Yun}
\email{jiyoungyun@kaist.ac.kr}
\affiliation{Information \& Electronics Research Institute, Korea Advanced Institute of Science and Technology (KAIST), $291$ Daehak-ro, Yuseong-gu, Daejeon $34141$, Republic of Korea }

\author{Seung-Hyun Nam}
\email{shnam@kaist.ac.kr} 
\affiliation{Information \& Electronics Research Institute, Korea Advanced Institute of Science and Technology (KAIST), $291$ Daehak-ro, Yuseong-gu, Daejeon $34141$, Republic of Korea }

\author{Hyun-Young Park}
\email{phy811@kaist.ac.kr}  
\affiliation{School of Electrical Engineering, Korea Advanced Institute of Science and Technology (KAIST), $291$ Daehak-ro, Yuseong-gu, Daejeon $34141$, Republic of Korea }

\author{Ashutosh Rai}
\email{ashutosh.rai@kaist.ac.kr}
\affiliation{School of Electrical Engineering, Korea Advanced Institute of Science and Technology (KAIST), $291$ Daehak-ro, Yuseong-gu, Daejeon $34141$, Republic of Korea }

\author{Si-Hyeon Lee}
\email{sihyeon@kaist.ac.kr} 
\affiliation{School of Electrical Engineering, Korea Advanced Institute of Science and Technology (KAIST), $291$ Daehak-ro, Yuseong-gu, Daejeon $34141$, Republic of Korea }

\author{Joonwoo Bae}
\email{joonwoo.bae@kaist.ac.kr}
\affiliation{School of Electrical Engineering, Korea Advanced Institute of Science and Technology (KAIST), $291$ Daehak-ro, Yuseong-gu, Daejeon $34141$, Republic of Korea }


\begin{abstract}

In this work, we consider two-sender, one-receiver communication over a discrete memoryless multiple-access channel without feedback, where two senders may cooperate on channel coding by using preshared resources, such as shared randomness, quantum states and measurements, or nonlocal correlations. We present the capacity region when senders employ cooperative encoding with quantum and nonlocal resources, extending beyond shared randomness, and derive a sum rate that serves as a lower bound to the sum capacity; the lower bound is computable by exploiting specific strategies. We also compute the sum capacities for two instances. One is when senders apply local resources for cooperative encoding. The other is when senders exploit nonclassical resources for encoding against channels constructed by referring to nonlocal games; in this way, correlated noise other than independent errors occurs on code words. Comparing the exact sum capacities and lower bounds, we show that nonlocal and quantum resources for cooperative encoding enable higher sum capacities over local ones. The Clauser-Horne-Shimony-Holt and magic square games are considered for constructing multiple-access channels, and we demonstrate the usefulness of nonlocal and quantum resources to achieve higher-sum capacities.
\end{abstract}

\maketitle
\section{Introduction}



Quantum advantages in communication signify the potential to surpass the limitations of current communication technologies by using quantum resources. The first attempt was shown in the seminal work on the channel capacity in point-to-point communication \cite{Holevo2011, 651037, PhysRevA.56.131}, where two parties distribute quantum states for the exchange of classical messages. The Holevo-Schumacher-Westmoreland theorem quantifies classical information that can be shared reliably between two parties sharing quantum states.

The usefulness of quantum resources over classical ones for point-to-point communication has been demonstrated as follows. Quantum key distribution \cite{RevModPhys.74.145} allows two honest parties to distribute quantum states and extract secret bits from measurement outcomes without the computational assumption. All prepare-and-measure protocols can be equivalently mapped into an entanglement-based scheme where two parties share entanglement \cite{PhysRevLett.68.557, PhysRevLett.92.217903, PhysRevLett.94.020501}. Entanglement-based realizations can be used to realize device-independent cryptographic protocols \cite{PhysRevLett.67.661, PhysRevLett.98.230501}. Moreover, a sender and a receiver sharing entanglement in advance can achieve higher classical channel capacity, known as entanglement-assisted channel capacities \cite{PhysRevLett.83.3081}. One-shot communication primitives, such as random access codes \cite{10.5555/795665.796491} and non-local games \cite{Brunner2014,Buhrman2010}, also achieve higher success probabilities with quantum resources than classical ones. 

In network information theory, multiple senders sharing entangled states or nonlocal correlations, but not with receivers, may achieve quantum advantages in channel capacities. The first instance is the case of two-sender and two-receiver interference channels \cite{QueckShor2017, YunRaiBae2020}, with senders sharing quantum and non-local correlations for encoding. Two senders sharing quantum and non-local correlations achieve higher channel capacities than classical ones. Then, higher channel capacities are also shown for two-sender and one-receiver multiple access channels \cite{Leditzky2020, Seshadri2023}. 

Higher channel capacities with quantum resources in a network are demonstrated by instances of network channels constructed using nonlocal games, such as the magic square game where quantum strategies win the game with certainty and the Clauser–Horne–Shimony–Holt (CHSH) game where nonlocal resources give the perfect strategy. Network channels in Refs. \cite{QueckShor2017, Leditzky2020} are constructed from the CHSH and the magic square games, respectively, such that they introduce noise only when codewords from multiple senders do not provide a perfect strategy; otherwise, messages are sent in a noiseless manner. They are also generalized with depolarization noise \cite{YunRaiBae2020, Leditzky2020, Seshadri2023}.  

It is, however, not yet straightforward to investigate other network channels that generally do not involve quantum or non-local games. What is lacking is the characterization of the capacity region for network channels when resources beyond classical ones are incorporated.

In this work, we consider two-sender and one-receiver multiple access channels (MACs) and present the framework for characterizing a capacity region when senders cooperate in encoding with quantum and non-local correlations. The framework rephrases a generalization of the capacity region for a MAC when two senders exploit non-classical correlations for channel coding. From the framework, we derive a sum rate in a single-letter formula that serves as a lower bound to the sum capacity, and the sum rate can be evaluated by exploiting specific strategies. 

We also present the capacity region and the sum capacity for MACs. When senders rely on local resources for cooperative encoding, we apply the Blahut-Arimoto algorithm \cite{Rezaeian_Grant_AB_algoMAC_2004, Watanabe_Kamoi_2002_AB_algoMAC_Article} to compute the sum capacity. We compute the sum capacities for MACs from quantum or nonlocal games containing various types of depolarization noise. We exploit the CHSH game and its depolarization noise to construct MACs and compute sum capacities with non-local and local encodings, as well as a lower bound to the quantum sum capacity. We also apply the magic square game to construct a MAC and compute the sum capacities. From MACs defined by referring to the quantum and nonlocal games, we demonstrate that nonlocal or quantum resources are strictly more useful than local ones for achieving higher channel capacities. 

Let us mention the technique in Refs. \cite{Leditzky2020, Seshadri2023}, which has derived an upper bound on the sum capacity. For instance, one can compute the upper bound to the sum capacity for a MAC when senders apply classical strategies, where quantum resources may achieve the maximal value of the capacity. Then, an upper bound can be used to demonstrate the existence of a separation between quantum and classical resources; quantum advantages are thus concluded. Our results provide the exact computation of the sum capacities here and also the capacity region, enabling direct comparisons of channel capacities. Our results span MACs for various types of depolarization noise with parameters in a wide range; hence, the comparisons of channel capacities, as well as the conclusions drawn from them, are valid for a larger class of MACs. Importantly, we establish an information-theoretic framework for characterizing the sum capacity, from which we also derive a sum rate in a single-letter formula that serves as a lower bound to the sum capacity. The single-letter characterization can generally be computed by taking a specific strategy.

This paper is organized as follows. In Sec. \ref{sec2}, we review the framework of a two-sender and one-receiver discrete memoryless MAC (DM-MAC) without feedback. We also summarize non-signaling correlations, such as local, quantum and non-local ones, as resources for cooperation by two senders. We also summarize the Blahut-Arimoto algorithm. In Sec. \ref{sec3}, we present a multi-letter characterization of the capacity region by incorporating the cooperation of two senders into channel coding. A sum rate that serves as a lower bound to the sum capacity is derived. In Sec. \ref{sec4}, we present the construction of MACs from quantum and nonlocal games and compute exact values of sum capacities when cooperative encoding applies a perfect winning strategy with quantum or nonlocal resources. Capacities with local strategies are strictly away from the maximal value. In Sec. \ref{sec5}, MACs from the CHSH game are provided and elucidate the advantages of quantum and nonlocal resources. In Sec. \ref{sec6}, MACs from the magic square game are provided and show the advantages of quantum resources over classical strategies. In Sec. \ref{sec7}, we conclude the results and discuss generalizations to other network channels. 
 
\section{Preliminaries: Multiple Access Channels}
\label{sec2}

In this section, we review MACs and summarize non-signaling correlations as resources for channel coding. Throughout this work, let $H(X)$ denote the entropy of a discrete random variable $X$ and  $H(Y|X)$ the conditional entropy of $Y$ given $X$. We also write by $H(X|Y=y)$ the conditional entropy of $X$ given an event $Y=y$. Let $I(X; Y)$ denote the mutual information between variables $X$ and $Y$, and $I(X; Y | Z)$ the conditional mutual information of $X$ and $Y$ given $Z$.
 
\begin{figure*}
    \centering
\includegraphics[width=0.8\linewidth]{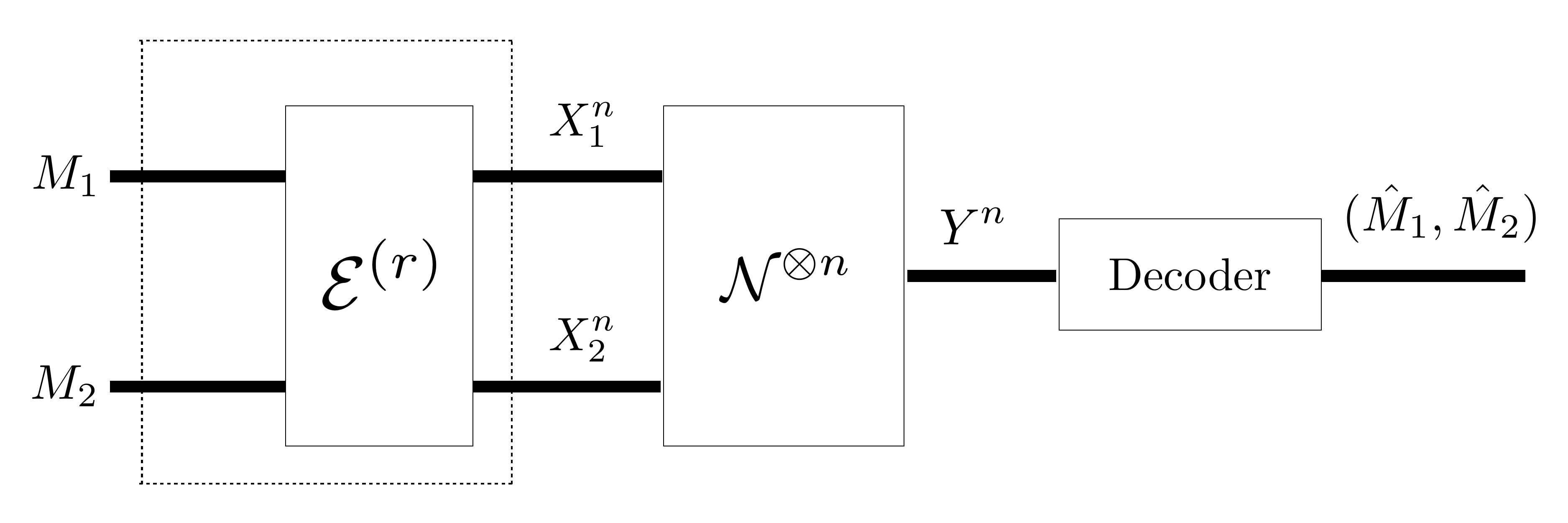}
    \caption{ A two-sender and one-receiver MAC is shown. The dotted box with cooperations $\E^{(r)}$ with resources $r\in \{ L,Q,NS\}$ shows the realization of cooperative encoding of senders. When the cooperation is compatible with local resources, the capacity region has been obtained in terms of the single-letter characterization \cite{Ahlswede,Liao}. The capacity region when senders exploit shared resources beyond classical ones is shown in Eq. (\ref{eq: C^r_mult}). }
    \label{fig:equiv_model}
\end{figure*}

We consider a communication scenario for two-sender and one-receiver discrete memoryless MAC (DM-MAC) without feedback, see Fig. \ref{fig:equiv_model}. In a discrete and memoryless channel,  all the input and output alphabets of the channel are finite and the outputs of a channel at any given time depend only on the current input, but not on previous inputs or outputs.
In this scenario, two senders aim to send messages $M_1$ and $M_2$ to a receiver.
We assume that messages $(M_1, M_2)$ are independent and also uniformly distributed over a finite set $\mathcal{M}_1 \times \mathcal{M}_2$.
For messages $(M_1,M_2)$, two senders prepare codewords $(X_1^n,X_2^n) \in \mathcal{X}_1^n \times \mathcal{X}_2^n$, by exploiting cooperation with shared resources, where we write by $r$ to denote shared resources.
Then, two senders transmit codewords $(X_1^n,X_2^n)$ over $n$ uses of a channel, i.e., $\mathcal{N}^{\otimes n}$ from $\mathcal{X}_1^n \times \mathcal{X}_2^n$ to $\mathcal{Y}^n$.
The receiver has a channel output $Y^n \in \mathcal{Y}^n$ and decodes it into $(\hat{M}_1,\hat{M}_2)$ by using a deterministic decoding function, called a decoder, $\mathcal{Y}^n \rightarrow \mathcal{M}_1 \times \mathcal{M}_2$.

\subsection{Encoding without shared resources}

When no shared resource is available, senders independently encode their messages $(M_1, M_2)$ into codewords $(X_1^n,X_2^n)$ by using deterministic encoding functions, called encoders,  $\mathcal{M}_1 \rightarrow \mathcal{X}_1^n$ and $\mathcal{M}_2 \rightarrow \mathcal{X}_2^n$.
The overall channel coding strategy is described by an $(n, R_1,R_2)$-code that consists of the followings:
    \begin{itemize}
        \item message sets $\mathcal{M}_i = \left[ 1:2^{ \lceil nR_i \rceil } \right]$ for $i = 1,2$,
        \item encoders $\mathcal{M}_i \rightarrow \mathcal{X}_i^n$ for $i = 1,2$, and a decoder $\mathcal{Y}^n \rightarrow \mathcal{M}_1 \times \mathcal{M}_2$.
    \end{itemize}
One can say that a rate tuple $(R_1,R_2)$ is achievable if there exists a sequence of $(n, R_1, R_2)$-codes for $n \in \mathbb{N}$ such that
    \begin{equation}\label{eq:no_error}
        \lim_{n\rightarrow \infty} \mathrm{Pr}( (M_1,M_2) \neq (\hat{M}_1,\hat{M}_2) ) = 0.
    \end{equation}
   The capacity region, denoted as $\mathcal{C}(\mathcal{N})$, is defined as the closure of the set of all achievable rate pairs,
    \begin{equation}
        \mathcal{C} (\mathcal{N}) = \mathrm{cl}(\left\{(R_1,R_2) \mid (R_1,R_2) ~\text{is} ~\text{achievable} \right\}),
    \end{equation}
    where $\mathrm{cl}$ denotes the closure.
The sum capacity is given by
    \begin{equation}
        \mathcal{C}_{s}(\mathcal{N}) = \max\limits_{(R_1,R_2) \in \mathcal{C}(\mathcal{N})} \left(R_1 + R_2\right).
    \end{equation}
Remarkably, the capacity region for a DM-MAC has been shown in terms of the \emph{single-letter characterization}.
\begin{thm}[\cite{Ahlswede,Liao}]\label{thm:C^X}
The capacity region $\mathcal{C}^{}(\mathcal{N})$ is given by the convex hull of the union of the following set
\bea
     \left\{ (R_1,R_2): \begin{matrix} R_1 \leq I(X_1;Y|X_2), \\ R_2 \leq I(X_2;Y|X_1), \\R_1+R_2 \leq I(X_1,X_2;Y)\end{matrix}         \right\}, \nonumber
\eea
over all $p(x_1)p(x_2)$. The sum capacity is given by 
    \begin{equation}\label{eq:C_s}
        \mathcal{C}^{}_s(\mathcal{N}) = \max\limits_{p(x_1)p(x_2)} I(X_1,X_2;Y)
    \end{equation}
where the maximization runs over independent probabilities.    
\end{thm}

\subsection{Resources for cooperative encoding }

Two senders may exploit shared resources $r$ to realize a cooperative encoding. 
A cooperative encoding with shared resources can be described by a conditional probability $\varepsilon (x_{1}^n, x_{2}^n | m_1,m_2 )$.
This can be identified by local, quantum, or non-local correlation.  

For convenience, we write finite sets of alphabets by $\mathcal{A}_1,\mathcal{A}_2,\mathcal{B}_1$ and $\mathcal{B}_2$. For elements $a_i\in\mathcal{A}_i$ and $b_i\in \mathcal{B}_i$, $i=1,2$, we say that a conditional probability $\varepsilon(a_2,b_2|a_1,b_1)$ contains local correlations if it admits a decomposition
    \bea
 \varepsilon(a_2,b_2|a_1,b_1) = 
        \sum_{\lambda \in \Lambda} p(\lambda) p(a_2 |a_1, \lambda) p(b_2 |b_1, \lambda) \label{eq:loc_coop}
    \eea
    with shared randomness $\lambda$. Otherwise, correlations are called non-local, which may be constrained by the non-signaling condition, 
    \bea
        \sum_{b_2 \in \mathcal{B}_2}\varepsilon(a_2,b_2|a_1,b_1) &=& p(a_2|a_1),~\mathrm{and} \nonumber\\
        \sum_{a_2 \in \mathcal{A}_2}\varepsilon(a_2,b_2|a_1,b_1) &=& p(b_2|b_1).\label{eq:NS_coop}
    \eea
Quantum resources are identified by joint probabilities having quantum realizations. A joint probability is compatible with quantum theory if there exists a state $\rho$ and measurements, described by a positive-operator-valued-measure (POVM), $\{\Pi^A_{a_2|a_1} \otimes \Pi^B_{b_2|b_1}\}$ such that 
    \bea
        \varepsilon(a_2,b_2|a_1,b_1) = \tr  [ \Pi^A_{a_2|a_1} \otimes \Pi^B_{b_2|b_1}  ~\rho ].\label{eq:Q_coop}
    \eea
Note that the quantum set is strictly larger than the local one, and the non-local correlations are a superset of the quantum ones. 

Throughout, we write by $L$, $Q$, and $NS$ to signify local, quantum, and non-signaling resources, respectively. Let 
\bea
r\in \{L,Q,NS\} \label{eq:r}
\eea
denote a specific resource among them. For instance, the set of cooperative encodings with quantum resources is denoted by $\E^{(Q)}$. From the hierarchy of joint probabilities, it is clear that 
\bea
\E^{(L)} \subsetneq \E^{(Q)} \subsetneq \E^{(NS)} \label{eq:hier}
\eea
in the sense that an encoding by local resources can be realized by quantum ones, and quantum ones by non-signaling ones, but not vice versa.

In a communication scenario where the senders use a cooperative encoding $\varepsilon(x_1^n,x_2^n|m_1,m_2) \in \mathcal{E}^{(r)}$, an $(n,R_1,R_2,r)$-code is defined analogously to an $(n,R_1,R_2)$-code, except that the deterministic encoders are replaced by a cooperative encoding $\varepsilon(x_1^n,x_2^n|m_1,m_2) \in \mathcal{E}^{(r)}$.
Accordingly, we say a rate pair $(R_1,R_2)$ is achievable if there is a sequence of $(n,R_1,R_2,r)$-codes such that Eq.~\eqref{eq:no_error} holds.
The capacity region $\mathcal{C}^{(r)}(\mathcal{N})$ is defined as the closure of the set of all achievable rate pairs $(R_1,R_2)$, and the sum capacity is $\mathcal{C}_s^{(r)}(\mathcal{N}) = \max_{(R_1,R_2) \in \mathcal{C}^{(r)}(\mathcal{N})}(R_1+R_2)$.

\subsection{ The generalized Blahut-Arimoto algorithm }
\label{subsec:ba}

The generalized Blahut-Arimoto algorithm \cite{Rezaeian_Grant_AB_algoMAC_2004, Watanabe_Kamoi_2002_AB_algoMAC_Article} provides a numerical approach to compute the sum capacity. The main idea is to reformulate the channel capacity, for instance for point-to-point communication, in terms of $p(x)$ and $p(y|x)$ as follows 
\bea
\max_{p(x)}  I (X;Y) =\max_{p(x), p(x|y)} \sum_{x,y} p(x)p( y|x) \log \frac{p( x|y)}{p(x)} \nonumber
\eea
where $p(y|x)$ is fixed from a channel, and the maximum runs over all probabilities $p(x)$, and all conditional probabilities $p(x|y)$ satisfying the condition $p(x|y)=0$ if and only if $p(y|x) = 0$ \cite[Thm.~10.2]{yeungFirstCourseInformation2002}. Then, an alternating maximization finds the channel-achieving distributions by iterative updates.

For MACs where senders exploit classical strategies, the sum capacity can be obtained by solving the maximization in Eq. (\ref{eq:C_s}). The Blahut-Arimoto algorithm can be generalized to two senders by reformulating the maximization in the sum capacity as,
\bea
&&  \max_{ p(x_1) p(x_2)}  I (X_1,X_2;Y) \nonumber\\
&=& \max_{ } \sum_{x,y} p(x_1)p(x_2) p(y|x_1,x_2) \log \frac{p( x_1,x_2|y)}{p(x_1) p(x_2)}. \nonumber
\eea
In the above, $p(y|x_1,x_2)$ is fixed from a MAC, and the maximum runs over probabilities $p(x_1)$, $p(x_2)$ and $q(x_1,x_2|y)$ satisfying the condition $q(x_1,x_2|y)=0$ if and only if $p(y|x_1,x_2) = 0$.

With an iterative alternating maximization approach, one starts with randomly chosen probabilities, $p^{(0)}(x_1)$ and $p^{(0)}(x_2)$, of two senders, respectively. Suppose that $|\mathcal{X}_1| , |\mathcal{X}_2| \leq |\mathcal{Y}|$. The probability $p^{(t+1)}(x_i)$ of the $i$-th sender   at the $(t+1)$-th iteration can be found as follows, 
\begin{equation}
    p^{(t+1)}(x_i) = p^{(t)}(x_i) \frac{\exp(I^{(t)}(x_i;Y))}{\sum\limits_{x'_i} p^{(t)}(x_i') \exp(I^{(t)}(x_i';Y))},
\end{equation}
where
\begin{align}
    I^{(t)}(x_i;Y) &= \sum\limits_{x_j} p^{(t)}(x_j) \sum\limits_y p(y|x_1,x_2) \log \frac{p(y|x_1,x_2)}{p^{(t)}(y)}, \nonumber
\end{align}
for $i \neq j$ and $p^{(t)}(y)$ is the marginal distribution of the channel output at the $t$-th iteration.
Then, the mutual information $I^{(t)}(X_1 , X_2 ; Y)$ converges to the sum capacity as $t$ tends to be large \cite[Thm.~2]{Rezaeian_Grant_AB_algoMAC_2004}.


We note that the generalized Blahut-Arimoto algorithm may not generally converge to the true sum capacity for arbitrary MACs \cite{Bhler2011ANO}. We then extensively explore all encodings to directly compute the sum capacity, via numerical searches of the maximum sum rate, and find that the extensive search reproduces the results of the generalized Blahut-Arimoto algorithm.

If $|\mathcal{X}_i| > |\mathcal{Y}|$ for some $i$, the algorithm applies to every regular sub-MACs, denoted by $\tilde{\mathcal{N}}$ of $\mathcal{N}$, in a parallel manner. A regular sub-MAC $\tilde{\mathcal{X}}_1 \times \tilde{\mathcal{X}}_2$ to $\mathcal{Y}$ can be obtained by discarding a fraction of the alphabet of senders for $\mathcal{N}$ such that $|\tilde{\mathcal{X}_1}|, |\tilde{\mathcal{X}_2}| \leq |\mathcal{Y}|$. Since a regular sub-MAC exists such that the sum capacity is equal to the original one \cite[Thm.~1]{Watanabe_Kamoi_2002_AB_algoMAC_Article}, the sum capacity of a MAC can be obtained by, at each iteration, taking the maximum over the resulting mutual information calculated for each regular sub-MAC.

\section{Capacity Regions: Cooperative encoding}
\label{sec3}

In this section, we present the capacity region of two-sender and one-receiver DM-MACs when senders exploit shared resources for cooperative encoding. The resources are denoted by $r$, see Eq. (\ref{eq:r}).

\subsection{ Capacity regions }

Firstly, two senders may share local resources that can be described by joint probabilities in Eq. (\ref{eq:loc_coop}). Since the set of local correlations is convex, the capacity region is determined by the extreme points of the convex set, at which local resources are characterized by independent probabilities. It follows that the capacity region with local resources is equal to instances where no shared resource is available. \\

\begin{prop}\label{prop:C^L}
For MACs, senders cannot improve the capacity region with local resources such as shared randomness. 
\end{prop}

Next, two senders may share nonclassical resources, such as quantum or non-local correlations, i.e., $r \in \{Q, NS\}$.
{\ar In this case, a multi-letter characterization of the capacity region with quantum resources, $\mathcal{C}^{(Q)}$, has been shown recently in Ref. \cite[Thm.~2]{pereg2025}. Since the proof in Ref. \cite[Thm.~2]{pereg2025} is valid for all $r \in \{L,Q,NS\}$, we apply the result to characterizing the capacity region with resources $r$ for all $r \in \{L,Q,NS\}$. Therefore, we present the characterization as follows. 
}


\begin{prop}\label{prop:MultiLetterFormula}
    When two senders share resources $r \in \{L,Q,NS\}$, the capacity region is given by
    \begin{equation}
    \mathcal{C}^{(r)} (\mathcal{N}) =   \mathrm{cl}\left(\bigcup\limits_{k \in \mathbb{N}}  \mathcal{C}^{(r),k}(\mathcal{N}) \right),\label{eq: C^r_mult}
    \end{equation}
    where 
    \begin{equation}
        \mathcal{C}^{(r),k}(\mathcal{N}):= \left\{ (R_1,R_2): \begin{matrix} R_1 \leq \frac{1}{k}I(U_1;Y^k|U_2), \\ R_2 \leq \frac{1}{k}I(U_2;Y^k|U_1), \\R_1+R_2 \leq \frac{1}{k} I(U_1,U_2;Y^k). \end{matrix} 
        \right\}, \nonumber
         \end{equation}
for some $p(u_1)p(u_2)$ and $\varepsilon(x_1^k,x_2^k|u_1,u_2) \in  \mathcal{E}^{(r)}$.
The sum capacity is given by
    \begin{multline}
        \mathcal{C}_s^{(r)}(\mathcal{N})
        \\ = \sup\limits_{k\in \mathbb{N}}  \sup\limits_{p(u_1)p(u_2)} \sup\limits_{\varepsilon(x_1^k,x_2^k|u_1,u_2) \in  \mathcal{E}^{(r)}} \frac{1}{k} I(U_1,U_2;Y^k). \label{eq:C_s^r_mult}
    \end{multline}
For cases with $ r=L$, the region above is equal to the single-letter characterization in Eq. (\ref{eq:C_s}) \cite{Ahlswede,Liao}. 
\end{prop}


{\ar Let us briefly review the achievability part of the proof of the proposition and also clarify the auxiliary random variables $U_1$ and $U_2$.
The key to derive the achievability is to consider deterministic local pre-encodings $\mathcal{M}_i \rightarrow \mathcal{U}_i^{n}$, $i = 1,2$, and a block-wise cooperative encoding $\varepsilon(x_1^k,x_2^k|u_1,u_2)$ over channel uses $\mathcal{N}^{\otimes k}$. One can apply the standard random coding for the effective channel, denoted by $\mathcal{N}^{\otimes k} \circ \varepsilon(y^k |u_1,u_2)$, to characterize an achievable rate region.

In more detail,} for each $k \in \mathbb{N}$, we construct a sequence of $(nk,R_1,R_2,r)$-codes for $n \in \mathbb{N}$.
In this construction, we fix distributions $p(u_1)$, $p(u_2)$, and $\varepsilon(x_1^k,x_2^k | u_1,u_2) \in \mathcal{E}^{(r)}$.
Then, for each $i=1,2$, and $m_i \in \mathcal{M}_i$, a pre-codeword $u_i^n(m_i)$ is generated according to $p(u_i)$ in an i.i.d. manner.
At the encoding stage, two senders first independently encode their messages $(M_1,M_2)$ into pre-codewords $(U_1^{n},U_2^{n})$ via pre-encoders $\mathcal{M}_1 \rightarrow \mathcal{U}_1^n$ and $\mathcal{M}_2 \rightarrow \mathcal{U}_2^n$.
Then, they generate codewords $(X_1^{nk},X_2^{nk})$ from $(U_1^{n},U_2^{n})$ by exploiting $\varepsilon^{\otimes n}(x_1^{nk},x_2^{nk} | u_1^n,u_2^n)$.
At the decoding stage, the receiver decodes $Y^{nk}$ into $(\hat{M}_1,\hat{M}_2)$ by using joint typicality decoding.
By viewing $\mathcal{N}^{\otimes k} \circ \varepsilon(y^k |u_1,u_2)$ as an effective channel and applying the standard random coding argument, we prove the achievability part.

We note that it remains challenging to derive a single-letter characterization of the capacity region \cite{Gamal_and_Kim}. 

\subsection{ Sum rate as a lower bound to sum capacities}

The multi-letter characterization in Proposition \ref{prop:MultiLetterFormula} allows us to derive inner bounds and to demonstrate strict inequalities among sum capacities. Let $\R^{(r)}(\mathcal{N})$ denote an inner bound of the capacity region for a channel $\N$ when a resource $r$ is exploited in encoding, as follows,
\begin{equation}
    \mathcal{R}^{(r)}(\mathcal{N}):=\mathcal{C}^{(r),1}(\mathcal{N}) \subseteq \mathcal{C}^r(\mathcal{N}). \nonumber
\end{equation}
{\ar The inner bound above $\mathcal{R}^{(r)}$ is the set of all achievable rate pairs when the senders impose restriction in encoding as follows. Senders apply channel-wise cooperative encoding described as $\varepsilon(x_1,x_2|u_1,u_2)$; in contrast to this, a block-wise encoding is denoted as, $\varepsilon(x_1^k,x_2^k|u_1,u_2)$ for $k \geq 2$.}
From the inner bound above, we derive an achievable sum rate in the single-letter characterization.

\begin{cor}\label{cor:compl}
The sum capacity in Eq. (\ref{eq:C_s^r_mult}) has a lower bound,
\bea
  \mathcal{C}_s^{(r)}(\mathcal{N})\geq \mathcal{R}_s^{(r)} (\mathcal{N}) = \sup\limits_{p(u_1)p(u_2)} \sup\limits_{\mathcal{E}^{(r)}} I(U_1,U_2;Y),~~~ \label{eq:coml}
\eea
where $\mathcal{R}_s^{(r)}$ is called a sum rate with resource $r$. 
\end{cor}

Note that a rate $\mathcal{R}_s^{(r)}$ does not contain maximization over copies $k$ and thus can be computed by addressing specific strategies. Thus, it serves as a lower bound to the sum capacity. When senders apply quantum encoding and a specific strategy, a lower bound to the sum rate in Eq. (\ref{eq:coml}) can be obtained. A strict separation between local and quantum cooperations can be obtained by showing that
\bea
\mathcal{C}_s^{(L)}(\mathcal{N}) < \mathcal{R}_s^{(Q)}(\mathcal{N}). \nonumber
\eea
The sum capacity when senders apply local resources for cooperative encoding can be computed from the single-letter characterization, see Eq. (\ref{eq:C_s}) \cite{Ahlswede,Liao}.

\section{ Sum Capacity of MACs }
\label{sec4}

In this section, we consider DM-MACs by referring to nonlocal and quantum games. Parties win a quantum game with certainty if they apply quantum strategies. For a nonlocal game, parties should employ a strategy utilizing nonlocal resources to win the game with certainty. We provide the construction of DM-MACs from quantum and nonlocal games such that channel noise contains correlations compatible with quantum or nonlocal resources rather than independent errors.

{\ar Let us clarify that the quantum or nonlocal games we refer to throughout consider a scenario of players who are spatially separate and cooperate via shared resources to compute a function against a referee. The scenario introduces communication complexity to determine the number of rounds of communication required to compute the function \cite{Buhrman2010}. Then, a game that defines such a function is called quantum pseudo-telepathy when parties sharing quantum resources have the perfect winning strategy ~\cite{GHZ, Mermin}. It is also known as a non-local game when parties sharing non-local resources have a perfect winning strategy, such as the CHSH game \cite{chsh, Brunner2014}.  }

{\ar Note that in the literature, quantum games have also been used in Ref. \cite{EWL1999}, where they signify a quantum analogy of the game theory in economics. They are also used in classical Bayesian or incomplete-information games in which players have conflicting interests and
seek equilibrium solutions when such games are extended to quantum resources \cite{Brandenburger_2016, Koniorczyk+2020, Auletta+2021}. These are not under our consideration. 
}

\subsection{MACs from two-party games}

\begin{defn}
    A finite two-party game $G$ is defined as a tuple of finite sets $(\mathcal{A}_{1},\mathcal{A}_{2},\mathcal{B}_{1},\mathcal{B}_{2},\mathcal{W})$ where $(\mathcal{A}_{1}, \mathcal{B}_{1})$ are the inputs given to the two parties, $(\mathcal{A}_{2}, \mathcal{B}_{2})$ are the respective outputs. The set $\mathcal{W} \subseteq \mathcal{A}_{1}\times\mathcal{A}_{2}\times\mathcal{B}_{1}\times\mathcal{B}_{2}$ denotes the winning conditions.
\end{defn}

\begin{figure}
    \centering
    \includegraphics[width=0.9\linewidth]{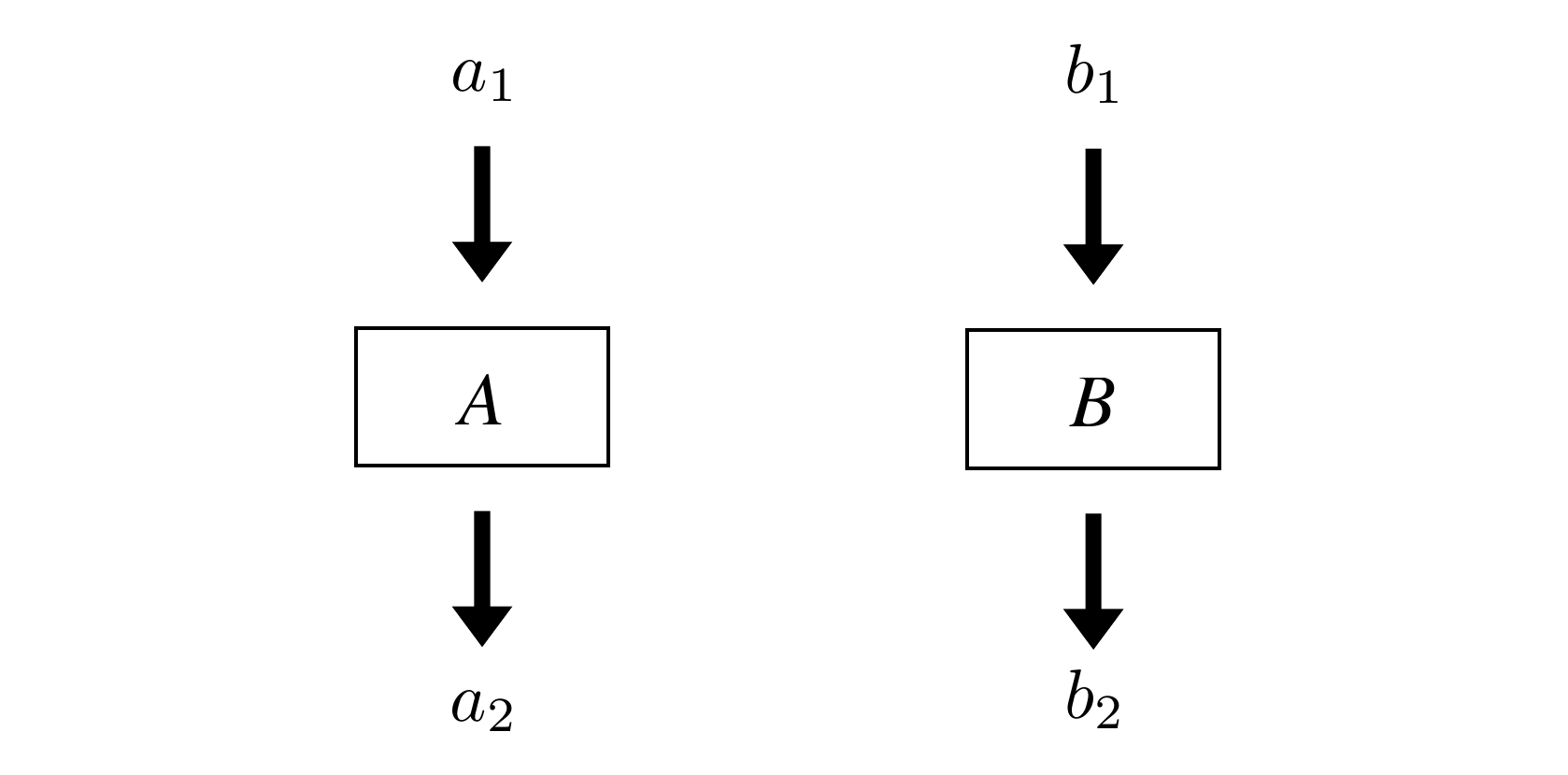}
    \caption{  Two parties $A$ and $B$ play a game $G$ against a referee who gives a query. Each party can select its own inputs and outputs. The parties are spatially separated and cannot communicate with each other. Two parties win the game if their inputs and outputs are in a set $\mathcal{W} \subset \mathcal{A}_{1}\times \mathcal{A}_{2} \times \mathcal{B}_{1} \times \mathcal{B}_{2} $ of winning strategies, i.e., $(a_{1}, a_{2}, b_{1}, b_{2})\in \mathcal{W}$. }
    \label{fig:nonlocalgame}
\end{figure}

Parties win a game if their inputs and outputs satisfy a winning condition, see Fig.~\ref{fig:nonlocalgame}. 

\begin{defn}
    Let $r \in \{Q,NS\}$ denote a resource. A game $G$ is called an $r$ pseudo-telepathy game, or simply an $r$ game, if there exists a perfect winning strategy with resources $r$. That is, for all input pairs $(a_1,b_1) \in \mathcal{A}_{1}\times \mathcal{B}_{1}$, parties win the game with certainty:
    \begin{equation}\label{eq:perf_win_cond}
        \mathrm{Pr}\big((A_1,A_2,B_1,B_2) \in \mathcal{W} | (A_1,B_1) = (a_1,b_1)\big) = 1,~
    \end{equation}
    where the conditional distribution of $(A_2,B_2)$ given $(A_1,B_1)$ follows from winning strategies $\varepsilon(a_2,b_2|a_1,b_1)$ with available resources $r$.
\end{defn}

We may refer to a quantum game when quantum resources provide a strategy to win a game with certainty, and also to a non-local game when non-local resources achieve a perfect strategy \cite{Brunner2014}. For instance, the CHSH game is a nonlocal one, as parties can win the game with certainty only when nonlocal resources are available. The magic square game is a quantum game since parties sharing quantum resources win the game with certainty \cite{Brassard2005}. In both cases, local strategies win the games with some probabilities; hence, the winning probability is strictly less than $1$. 


Network channels can be constructed from two-party games as follows. 

\begin{defn}
    From a game $G = (\mathcal{A}_{1},\mathcal{A}_{2},\mathcal{B}_{1},\mathcal{B}_{2},\mathcal{W})$, a MAC can be constructed such that
    \bea
    \mathcal{N}: \mathcal{X}_1 \times \mathcal{X}_2 \rightarrow \mathcal{Y} \nonumber
    \eea
    where $\mathcal{X}_1 = \mathcal{A}_{1} \times \mathcal{A}_{2}$, $\mathcal{X}_2 = \mathcal{B}_{1} \times \mathcal{B}_{2}$, and a finite set $\mathcal{Y}$. Then, a channel produces outcomes from input-output pairs of both parties playing a game $G$ containing a winning condition $\mathcal{W}$. For quantum or nonlocal games, a MAC introduces noise which are compatible with quantum or nonlocal correlations. 
\end{defn}

In the following, we present an upper bound on mutual information $I(X_1, X_2 ; Y)$ for a MAC defined by a game. We may use a notation $h(x_1,x_2)$ as an estimation of the uncertainty about $Y$ given $(X_1,X_2) =(x_1,x_2)$ in terms of the binary entropy,
\bea
h (x_1,x_2) & = &  H( Y | X_1=x_1, X_2=x_2 ). \nonumber 
\eea
In addition, we also write $h_m: = \min_{(x_1,x_2)} h(x_1,x_2)$ as the minimum value. 

\begin{lem}\label{lem:game_ch_Ixy}
   Consider a MAC $\mathcal{N}: \mathcal{X}_1 \times \mathcal{X}_2 \rightarrow \mathcal{Y}$ defined by a game $G$. For the inputs and outcomes of the channel, it holds that
    \begin{align}
        I(X_1,X_2;Y) 
        & \leq  \log |\mathcal{Y}| - h_m. \label{eq:lem1_ineq}
    \end{align}
Note that the upper bound is valid for all input distributions $p(x_1,x_2)$. 
\end{lem}
\begin{proof}
For simplicity, let $Z = (X_1,X_2)$. Then, the mutual information can be expressed as follows,
\bea
  	I(Z;Y) &=& H(Y) - H(Y\mid Z)  \nonumber\\
  	&=& H(Y) - \sum_{z}p(z)H(Y\mid Z=z). \nonumber
\eea	
Note that $H(Y) \leq \log |\mathcal{Y}|$, and we have
\bea
\sum_{z}p(z)H(Y\mid Z=z) &\geq & \sum_{z}p(z) \min_{z} H(Y\mid Z=z) \nonumber \\
& = &  \min_{z} H(Y\mid Z=z). \nonumber
\eea
The inequality in Eq. \eqref{eq:lem1_ineq} is thus shown.
\end{proof}

We exploit Lemma~\ref{lem:game_ch_Ixy} to derive a general upper bound on the sum capacity of the channel for different types of encoding strategies. 

\begin{lem}\label{lem:C_s^r_UB}
    For a resource $r \in \{L,Q,NS\}$ and a MAC $\mathcal{N}$ defined by a game $G$, it holds that 
	\begin{align}
	  \mathcal{C}_s^{(r)} \left(\mathcal{N}\right) \leq  \log |\mathcal{Y}| -    h_m. \label{eq:up}
	\end{align}   
This shows a general upper bound to the sum capacity, see Eq. (\ref{eq:C_s^r_mult}), and the bound is valid for cooperative encoding with resources $r$. 
\end{lem}
\begin{proof}
    By Proposition~\ref{prop:MultiLetterFormula}, it is sufficient to show that for any $n \in \mathbb{N}$, finite sets $\mathcal{U}_1$, $\mathcal{U}_2$, $p(u_1)p(u_2)$, and $\varepsilon(u_1,u_2|x_1^n,x_2^n) \in \mathcal{E}^{(r)}$, we have 
    \begin{align}
        \frac{1}{n}I(U_1, U_2;Y^n) \leq \log |\mathcal{Y}| - h_m.
    \end{align}
    By the data processing inequality and the memoryless property of the DM-MAC, we obtain
    \begin{align}
        \frac{1}{n}I(U_1,U_2;Y^n) &\leq \frac{1}{n} I(X_1^n,X_2^n;Y^n)
        \\& \leq \frac{1}{n}\sum\limits_{j=1}^n I(X_{1j},X_{2j};Y_j).
    \end{align}
    We apply Lemma~\ref{lem:game_ch_Ixy} and find the upper bound to the sum capacity in Eq. (\ref{eq:up}). 
\end{proof}

We remark that the two lemmas hold for general MACs that may not be constructed by referring to games.

With upper and lower bounds on sum capacities in Eqs. (\ref{eq:coml}) and (\ref{eq:up}), we consider MACs constructed from $r$ pseudo-telepathy games and also their generalizations including noise parameters. Various types of depolarization noise can be considered as follows. 

\begin{defn}
    Consider a game $G = (\mathcal{A}_{1},\mathcal{A}_{2},\mathcal{B}_{1},\mathcal{B}_{2},\mathcal{W})$ and noise parameters, denoted by $\eta_w$ and $\eta_l$, such that $1\geq  \eta_w > \eta_l\geq 0$. From the game, we construct a $(G,\eta_w,\eta_l)$-depolarizing MAC, 
\bea
\mathcal{N}_G^{(\eta_w,\eta_l)}: \mathcal{X}_1\times \mathcal{X}_2 \rightarrow \mathcal{Y}, \label{eq:nn}
\eea    
where $\mathcal{X}_1 = \mathcal{A}_{1} \times \mathcal{A}_{2}$, $\mathcal{X}_2 = \mathcal{B}_{1} \times \mathcal{B}_{2}$, and $  \mathcal{Y} = \mathcal{A}_{1} \times \mathcal{B}_{1}$, and the channel is characterized by a conditional probability $ p_{G}^{ (\eta_w,\eta_l) } (y|x_1,x_2)$ where
    \begin{multline}
        \!\!\!  p_{G}^{ (\eta_w,\eta_l) } ( y_1,y_2|a_1,a_2,b_1,b_2)\nonumber
        \\\!\! = \begin{cases}
 			\eta_w ~ \delta_{y_1,a_1}\delta_{y_2,b_1} \!\!+\! (1-\eta_w)\frac{1}{|\mathcal{Y}|} ,& \!\!\! \mbox{if}~ (a_1,a_2,b_1,b_2)\in\mathcal{W} \\  
 			\eta_l ~ \delta_{y_1,a_1}\delta_{y_2,b_1} \!\!+\! (1-\eta_l) \frac{1}{|\mathcal{Y}|},& \!\!\! \mbox{otherwise.}
 		\end{cases}
    \end{multline}
Note that the probability $p(b|a) = \delta_{b,a}$ characterizes an identity map, $\mathrm{id}[a]=  a$ for an input $a$.
\end{defn}

Note that in terms of the parameters of a MAC above, one can have the MAC in Ref. \cite{Leditzky2020} as an instance, which is a $(G,1,0)$-depolarizing game channel. The channel introduces a noiseless channel when cooperative encoding follows from a perfect winning strategy; otherwise, it results in complete depolarization. MACs with two parameters $(\eta_w,\eta_l)$ show a generalization of the MAC in Ref. \cite{Leditzky2020}; we introduce depolarization noise for both cases over a wide range.

\subsection{ Sum capacity of MACs from two-party games}

For $(G,\eta_w,\eta_l)$-depolarizing MACs defined in Eq. (\ref{eq:nn}), we here compute the exact sum capacity. For convenience, let $h(\eta)$ denote a binary entropy with a parameter $\eta$ as follows, 
\bea
&&    h(\eta) = - \alpha(\eta)\log \alpha(\eta) - (|\mathcal{Y}|-1) \beta(\eta)\log \beta(\eta)~~~  \label{eq:hh}\\
   && \mathrm{where} ~~ \alpha(\eta) = \frac{1+(|\mathcal{Y}|-1)\eta}{|\mathcal{Y}|}~\mathrm{and}~\beta(\eta)= \frac{1-\eta}{|\mathcal{Y}|}. \nonumber
\eea
and $\mathcal{Y}$ is the set of alphabets of a receiver. Note that $h(\eta)$ may also be derived as the entropy of outcomes of channels in Eq. (\ref{eq:nn}). 


\begin{thm}\label{thm:main}
Let $\mathcal{N}_G^{(\eta_w,\eta_l)}$ denote a $(G,\eta_w,\eta_l)$-depolarizing MAC for a game $G$ with noise parameters $\eta_w$ and $\eta_l$. For resources $r \in \{Q,NS\}$, if $G$ is an $r$ pseudo-telepathy game, i.e., strategies with $r$ resources win the game with certainty, then the sum capacity of the channel is given by,
    \begin{equation}
\mathcal{C}^{(r)}_s\left( \mathcal{N}_G^{(\eta_w,\eta_l)} \right) = \log|\mathcal{Y}| -h(\eta_w) \label{eq:capa}
    \end{equation}
where $h(\cdot)$ is in Eq. (\ref{eq:hh}). In addition, if local resources cannot win the game $G$ with certainty, then cooperative encoding with local strategies cannot achieve the maximal sum capacity, i.e.,
    \begin{equation}    \mathcal{C}^{(L)}_s(\mathcal{N}_G^{(\eta_w,\eta_l)}) < \log |\mathcal{Y}| - h(\eta_w). \label{eq:locallower}
    \end{equation}
Therefore, $r$ pseudo-telepathy games can be used to construct MACs to show the advantage of $r$ resources over local ones.  
\end{thm}

The result above shows that one can compute the sum capacity of MACs constructed from a game when senders employ cooperative encoding based on winning strategies. The usefulness of nonlocal and quantum resources becomes apparent in games where two parties cannot win with certainty using local resources.

\begin{proof}
From Lemma~\ref{lem:C_s^r_UB}, we have the upper bound
\begin{align}
  \mathcal{C}_s^{(r)} \left(\mathcal{N}\right) \leq  \log |\mathcal{Y}| -    h_m. \nonumber
\end{align}   
Moreover, the sum capacity is bounded below by a sum rate in Eq. (\ref{eq:coml})   
\bea
\mathcal{C}_s^{(r)} \left( \mathcal{N}_G^{(\eta_w,\eta_l)} \right)\geq     \mathcal{R}_s^{(r)} \left( \mathcal{N}_G^{(\eta_w,\eta_l)} \right). \nonumber       
\eea
Then, it suffices to show that the following inequality also holds true,
    \begin{eqnarray}
        \mathcal{R}^{(r)}_s \left( \mathcal{N}_G^{(\eta_w,\eta_l)} \right) \geq \log|\mathcal{Y}| - h(\eta_w).\label{eq:lu}
    \end{eqnarray} 
To compute the sum rate above, we define $\mathcal{U}_1 = \mathcal{A}_{1}$, $\mathcal{U}_2 = \mathcal{B}_{1}$, and let $(U_1,U_2)\sim p(u_1)p(u_2)$, where $p(u_1)$ and $p(u_2)$ are uniform distributions over $\mathcal{U}_1$ and $\mathcal{U}_2$, respectively.
    Then, by Eq. (\ref{eq:coml}), we obtain
    \begin{eqnarray}
        \mathcal{R}^{(r)}_s \left( \mathcal{N}_G^{(\eta_w,\eta_l)} \right) \geq \sup\limits_{ \mathcal{E}^{(r)}} I(U_1,U_2;Y).
    \end{eqnarray}
    Now, let $\varepsilon'(a_{2},b_{2}|a_1,b_1)$ be a perfect winning strategy for the $r$ pseudo-telepathy game $G$, and 
    \begin{align}
    \varepsilon(a_{1},a_{2},b_{1},b_{2}|u_1,u_2) = \delta_{a_{1},u_1}\delta_{b_{1},u_2} \varepsilon'(a_{2},b_{2}|u_1,u_2).
    \end{align}
It follows that
    \bea
&&        \varepsilon \circ \mathcal{N}_G^{(\eta_w,\eta_l)} (y_1,y_2|u_1,u_2) = \eta_w~ \delta_{y_{1},u_1}\delta_{y_{2},u_2} \!\!+\! (1-\eta_w)\frac{1}{|\mathcal{Y}|}\nonumber
\eea
and we have the conditional entropy 
\bea
H(Y|U_1,U_2) = h(\eta_w).\nonumber
\eea
Since outcomes $Y$ follow a uniform distribution over the set $\mathcal{Y} = \mathcal{U}_1\times \mathcal{U}_2$, we conclude
    \begin{align}
        \mathcal{R}^{(r)}_s \left( \mathcal{N}_G^{(\eta_w,\eta_l)} \right) &\geq I(U_1,U_2;Y)
        \\& = H(Y) - H(Y|U_1,U_2)
        \\& = \log|\mathcal{Y}| - h(\eta_w).
    \end{align}
Thus, we have shown Eq. (\ref{eq:lu}), and computed the sum capacity in Eq. (\ref{eq:capa}).

    To show that local strategies do not achieve the maximal value of the sum capacity, see Eq. (\ref{eq:locallower}), we have the following, from Theorem~\ref{thm:C^X}, Proposition~\ref{prop:C^L}, and Lemma~\ref{lem:game_ch_Ixy},
    \begin{align}
        & \mathcal{C}_s^{(L)} \left( \mathcal{N}_G^{(\eta_w,\eta_l)} \right) = \max\limits_{p(x_1)p(x_2)} I(X_1,X_2;Y) \nonumber
        \\& = \max\limits_{p(x_1)p(x_2)} \left\{ H(Y) - h(\eta_l) + \varphi(h(\eta_l)-h(\eta_w)) \right\},\nonumber
    \end{align}
    where $\varphi = \mathrm{Pr}\big((X_1,X_2) \in \mathcal{W}\big)$.
    From Lemma~\ref{lem:C_s^r_UB}, suppose that local strategies achieve the maximal value of the sum capacity, that is, 
     \bea
    \mathcal{C}_s^{(L)}  \left( \mathcal{N}_G^{(\eta_w,\eta_l)} \right) = \log|\mathcal{Y}| - h(\eta_w). \label{eq:cont}
    \eea
This means $h_m= h(\eta_w)$, which holds if and only if there exist input distributions $p(x_1)$, $p(x_2)$ such that: (i) $\varphi = 1$, and (ii) the output distribution $p(y)$ is uniform over $\mathcal{Y}$. In the following, we show that if $\varphi=1$, then $p(y)$ cannot be uniform over $\mathcal{Y}$. 

Consider any arbitrary input distributions $p(x_1)$ and $p(x_2)$, and let $(X_1,X_2) \sim p(x_1)p(x_2)$, where we set $X_1 = (A_1,A_2)$, $X_2 = (B_1,B_2)$. Suppose that $\varphi = 1$. Then, there exists some $a_{1}^* \in \mathcal{A}_{1}$ such that $p(a_{1}^*)=\mathrm{Pr}(A_{1}=a_{1}^*)=0$. It follows that for all $b_{1} \in \mathcal{B}_{1}$
    \begin{align}
        &\mathrm{Pr}(Y=(a_{1}^*,b_{1})) = \frac{1-\eta_w}{|\mathcal{Y}|}<\frac{1}{|\mathcal{Y}|}.
    \end{align}
    This contradicts the assumption that $p(y)$ is uniform over $\mathcal{Y}$. Thus, we only consider the case that $p(a_1)=\mathrm{Pr}(A_1=a_1)>0$ and $p(b_1)=\mathrm{Pr}(B_1=b_1)>0$ for all $a_1 \in \mathcal{A}_1$ and $b_1 \in \mathcal{B}_1$.
    In this case, the conditional distributions $p(a_{2}|a_{1})$ and $p(b_{2}|b_{1})$ are well-defined.

    Since $p(a_{2}|a_{1})p(b_{2}|b_{1})$ is a local strategy that cannot achieve the perfect win in the game $G$, there exists $(a_{1}', a_{2}',b_{1}',b_{2}') \not\in \mathcal{W}$ such that
    \begin{equation}\label{eq:non_pseudo}
        p(a_{2}'|a_{1}')p(b_{2}'|b_{1}') > 0.
    \end{equation}
    Since $p(a_{1}')p(b_{1}')>0$, this implies that $\varphi < 1$, contradicting our assumption that $\varphi =1$. Hence, we do not achieve Eq. (\ref{eq:cont}) and have the strict inequality in Eq. (\ref{eq:up}) in Lemma~\ref{lem:C_s^r_UB}.
\end{proof}

We emphasize that Theorem~\ref{thm:main} implies that for any MAC containing depolarization on non-local or quantum games, see Eq. (\ref{eq:nn}), nonlocal or quantum cooperations of senders are more useful than local ones. In what follows, we consider MACs from games and show that sum capacities with nonlocal or quantum resources are strictly higher than local ones. We refer to nonlocal advantages if a higher channel capacity is achieved with nonlocal resources, i.e., $\mathcal{C}_s^{(NS)} > \mathcal{C}_s^{(L)}$, and quantum advantages when a higher channel capacity is obtained with quantum resources, i.e., $\mathcal{C}_s^{(Q)} > \mathcal{C}_s^{(L)}$.

    \begin{figure*}[t]
  	\centering
  	\begin{subfigure}[b]{0.495\textwidth}
  		\centering
  		\includegraphics[scale=0.25]{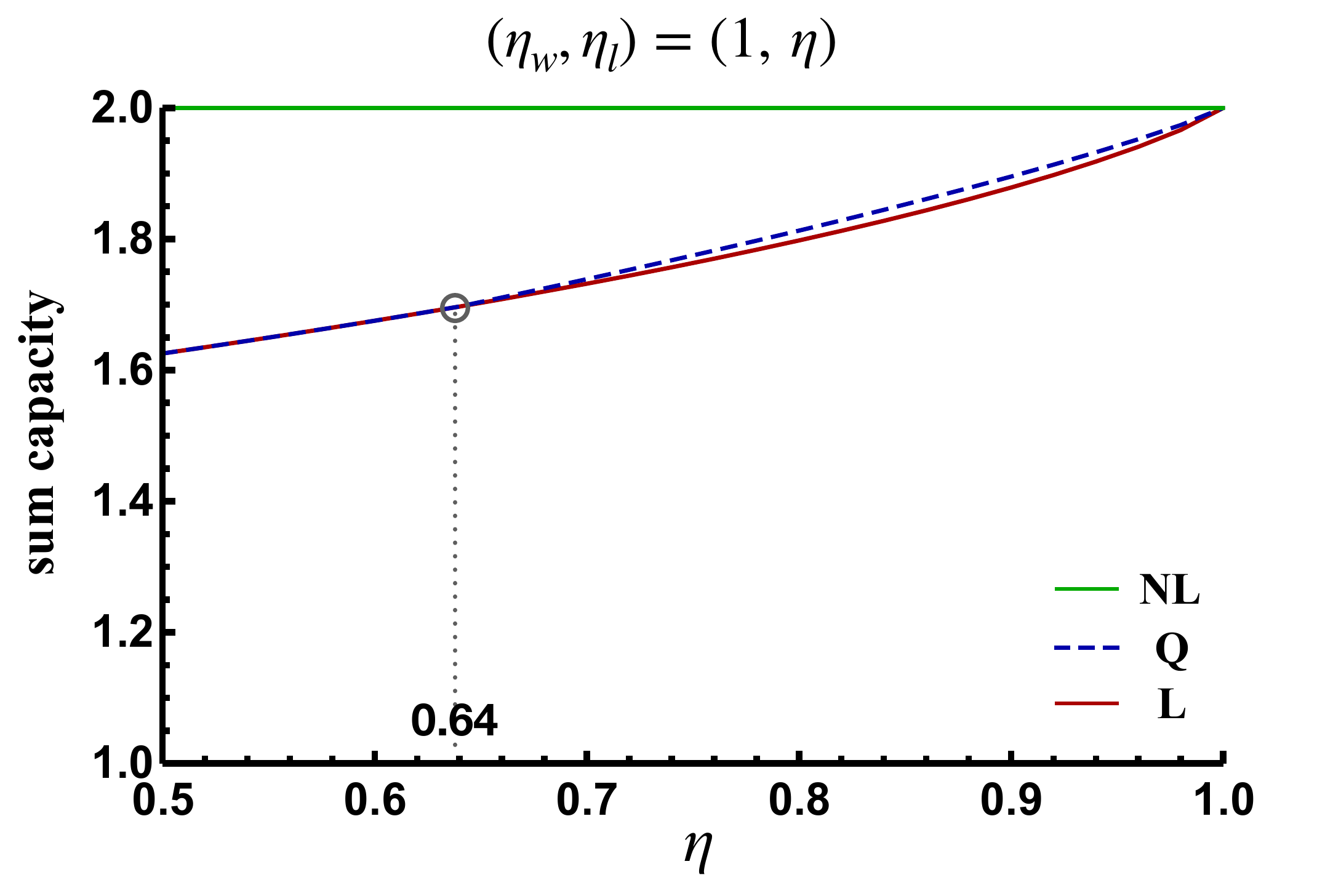}
  	\end{subfigure}
  	\hfill
  	\begin{subfigure}[b]{0.495\textwidth}
  		\centering
  		\includegraphics[scale=0.25]{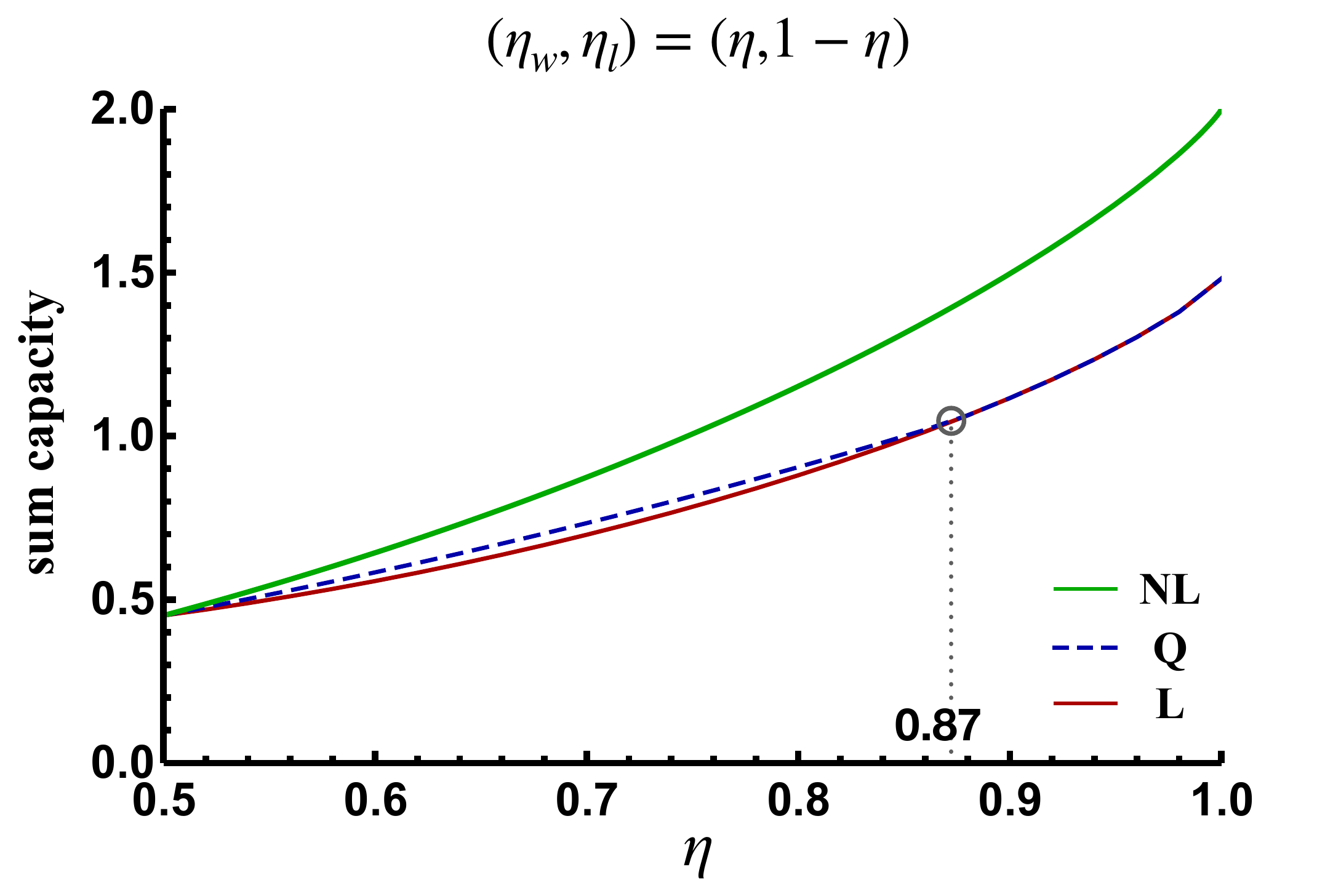}
  	\end{subfigure}
  	\caption{ For MACs constructed from the CHSH game with noise parameters $\eta_w$ and $\eta_l$, see Eq. (\ref{eq:nn}), sum capacities are computed. Two types of channel noise are considered for $(\eta_w,\eta_l) = (1,\eta)$ (left) and $(\eta_w,\eta_l) = (\eta,1-\eta)$ (right). The capacity when senders apply local resources (L) is obtained by using the Blahu-Arimoto algorithm (solid line in purple). Since it is a nonlocal game, the maximal value of the sum capacity can be achieved by nonlocal strategies. The sum capacity achieved through cooperative encoding with nonlocal resources is shown in Eq. (\ref{eq:nlchsh}) (green-colored solid line). Since quantum strategies do not win the game with certainty, a specific quantum strategy may serve as a lower bound (dashed line). For the type of noise $(\eta_w,\eta_l) = (1,\eta)$ (left), a lower bound to the sum capacity with quantum resources is higher than the classical capacity for $\eta \gtrsim 0.64$, while nonlocal encoding finds noiseless transmission. For the other case $(\eta_w,\eta_l) = (\eta, 1-\eta)$ (right), the sum capacity with nonlocal resources is higher than that with local ones, where a lower bound of the quantum capacity is higher than the classical one for $0.50<\eta \lesssim 0.87$. Both cases demonstrate the usefulness of quantum and nonlocal resources over local ones in achieving a higher sum capacity. }\label{fig:plot}
  \end{figure*}

\begin{table}[h!]
\centering
\begin{tabular}{cccc}
\toprule
$\eta$  ~~&~~ $C_s^{(L)}$  ~~&~~ $\mathcal{L}_s^{(Q)}$  ~~&~~ $C_s^{(NS)}$ \\[2pt]
\midrule
0.50 ~~&~~ 1.625  ~~&~~ 1.625  ~~&~~ 2.00 \\[2pt]
0.60  ~~&~~ 1.675  ~~&~~ 1.675  ~~&~~ 2.00 \\[2pt]
\rowcolor{gray!25}
0.64 ~~&~~ 1.6970  ~~&~~ 1.6971  ~~&~~ 2.00 \\[2pt]
0.70  ~~&~~ 1.732  ~~&~~ 1.739  ~~&~~ 2.00 \\[2pt]
0.80  ~~&~~ 1.798  ~~&~~ 1.813  ~~&~~ 2.00 \\[2pt]
0.90 ~~&~~ 1.879  ~~&~~ 1.895  ~~&~~ 2.00 \\[2pt]
1.00  ~~&~~ 2.00  ~~&~~ 2.00  ~~&~~ 2.00 \\[2pt]
\bottomrule
\end{tabular}
\caption{ For a MAC from the CHSH game with parameters $(\eta_w,\eta_l) = (1,\eta)$, see the subsection \ref{subsec:lq}. The sum capacity is computed when senders are assisted by resources: $\mathcal{C}_s^{(L)}$ and $\mathcal{C}_s^{(NS)}$ are capacities with local and nonlocal resources, respectively. For quantum resources, a lower bound, denoted by $\mathcal{L}_s^{(Q)}$ in Eq. (\ref{eq:lb}), to the quantum sum rate by addressing a specific strategy is shown. It is larger than the local one for $\eta \gtrsim 0.64$, for which quantum advantage is concluded.}
\label{tab:1}
\end{table}

\begin{table}[h!]
\centering
\begin{tabular}{cccc}
\toprule
$\eta$  ~~&~~ $C_s^{(L)}$  ~~&~~ $\mathcal{L}_s^{(Q)}$  ~~&~~ $C_s^{(NS)}$ \\[2pt]
\midrule
0.50  ~~&~~ 0.451  ~~&~~ 0.451  ~~&~~ 1.00 \\[2pt]
0.60  ~~&~~ 0.557  ~~&~~ 0.583  ~~&~~ 1.029 \\[2pt]
0.70  ~~&~~ 0.698  ~~&~~ 0.734  ~~&~~ 1.119 \\
0.80  ~~&~~ 0.880  ~~&~~ 0.905  ~~&~~ 1.278 \\[2pt]
\rowcolor{gray!25}
0.87 ~~&~~ 1.038   ~~&~~ 1.039   ~~&~~ 1.442 \\[2pt]
0.90  ~~&~~ 1.116   ~~&~~ 1.116  ~~&~~ 1.531 \\[2pt]
1.00  ~~&~~ 1.482   ~~&~~ 1.482   ~~&~~ 2.00 \\[2pt]
\bottomrule
\end{tabular}
\caption{ The sum capacity is computed for a MAC with parameters $(\eta_w,\eta_l) = (\eta,1-\eta)$ in the subsection \ref{subsec:lq}. For all ranges of $\eta$, the non-signaling sum capacity is higher than the local one. For $0.50 < \eta\lesssim 0.87$, quantum resources lead to a higher capacity since a lower bound $\mathcal{L}_s^{(Q)}$ in Eq. (\ref{eq:lb}) to the quantum sum capacity is strictly larger than the local one; hence, we have $\mathcal{L}_s^{(Q)}>\mathcal{C}_s^{(L)}$, for $0.50<\eta\lesssim 0.87$; quantum advantage is concluded. 
}
\label{tab:2}
\end{table}

\section{ Nonlocal and Quantum Advantages } 
\label{sec5}

One can define the CHSH game by a tuple $G_{CHSH}=(\mathcal{A}_{1},\mathcal{A}_{2},\mathcal{B}_{1},\mathcal{B}_{2},\mathcal{W})$ where $\mathcal{A}_{1} = \mathcal{A}_{2} = \mathcal{B}_{1} = \mathcal{B}_{2} = \{0,1\}$ and the winning strategy is
\bea
 \mathcal{W} = \{(a_1,a_2,b_1,b_2):a_2\oplus b_2 = a_1 b_1\}.\label{eq:chshwin}
\eea
From the game, $(G_{CHSH},\eta_w,\eta_l)$-depolarizing game channels can be constructed as it is shown in Eq. (\ref{eq:nn}). Let $\mathcal{N}_{CHSH}^{(\eta_w,\eta_l)}$ denote a MAC defined by referring to the CHSH game.

In the CHSH game, one can see the hierarchy of winning probabilities according to resources. Let $p_{r}^{(win)}$ denote a maximal winning probability with resources $r$; it holds that 
\bea
\frac{3}{4} = p_{L}^{(win)} < p_{Q}^{(win)} < p_{NL}^{(win)} =1. \nonumber
\eea
The CHSH game is a game where nonlocal strategies win with certainty. However, quantum strategies win the game with some probability, up to $(1/2) +(1/2\sqrt{2}) \approx 0.85$. Local strategies have a winning probability $3/4$.

\subsection{Nonlocal advantage}

The results in Theorem~\ref{thm:main} apply to MACs from the CHSH game with noise parameters $(\eta_w, \eta_l)$, see the construction of channels in Eq (\ref{eq:nn}). Firstly, we have the sum capacity for senders with nonlocal cooperation, 
\bea 
\mathcal{C}^{(NS)}_s\left(\mathcal{N}_{CHSH}^{(\eta_w,\eta_l)}\right)= 2 - h (\eta_w). \label{eq:nlchsh}
\eea 
Secondly, one can also find that the sum capacity with local strategies is strictly less than the maximal value,
\bea
\mathcal{C}^{(L)}_s\left(\mathcal{N}_{CHSH}^{(\eta_w,\eta_l)}\right) <  2 - h (\eta_w). \label{eq:lchsh} 
\eea
Note the results above hold for all $1\geq \eta_w > \eta_l\geq 0$. Hence, the usefulness of nonlocal resources over local ones in the channel capacities of MACs is shown.

\subsection{ Computing sum capacities: quantum vs. local strategies} 
\label{subsec:lq}

To show that quantum resources are more useful than classical ones, we consider two instances of depolarization parameters, see Tables~\ref{tab:1} and \ref{tab:2}.
\begin{itemize}
\item $(\eta_w,\eta_l)=(1,\eta)$, and
\item $(\eta_w,\eta_l)=(\eta,1-\eta)$.
\end{itemize}
The first case shows a noiseless channel when cooperative encoding follows from winning strategies; otherwise, depolarization noise may be present. The second case introduces noise with probability $\eta$ when two senders apply cooperative encoding from winning strategies; otherwise, noise appears with probability $1-\eta$.

As for sum capacities with local resources, one can apply the generalized Blahut-Arimoto algorithm \cite{Rezaeian_Grant_AB_algoMAC_2004, Watanabe_Kamoi_2002_AB_algoMAC_Article}, see the subsection \ref{subsec:ba}, and compute the sum capacity numerically. The results are shown in Fig.~\ref{fig:plot}.

We exploit a sum rate in Eq. (\ref{eq:coml}) for the sum capacity with quantum resources; the lower bound is referred to as a quantum achievable sum rate. It can be computed by applying a specific strategy as follows. Two senders share maximally entangled states $ |\Phi^{+}\rangle=\frac{1}{\sqrt{2}}(\vert00\rangle +\vert11\rangle)$. Before cooperative encoding with quantum resources, each party independently encodes messages: for parties $i=1,2$, $m_i$ into two bits $u_i=(u_{i1},u_{i2})$. Two parties computes $a_1=u_{11}\oplus u_{12}$ and $b_1=u_{21}\oplus u_{22}$. The cooperative encoding works by performing measurements
\bea
\{\Pi^A_{a_2|a_1}\}_{a_2 \in \{0,1\}} ~\mathrm{and}~ \{\Pi^B_{b_2|b_1}\}_{b_2 \in \{0,1\}} 
\eea 
on shared states $\ket{\Phi^+}^{}$. These are measurements to achieve the maximal violation of the CHSH inequality: $\Pi^A_{\cdot|0}$ and $\Pi^A_{\cdot|1}$ of the sender $A$ are, respectively, the observables $\sigma_z$ and $\sigma_x$, while the sender $B$ measurements $\Pi^B_{\cdot|0}$ and $\Pi^B_{\cdot|1}$ correspond to the respective observables $-(\sigma_z+\sigma_x)/\sqrt{2}$ and $(-\sigma_z+\sigma_x)/\sqrt{2}$. 

Then, resulting two-bit pairs $(a_1,a_2)$ and $(b_1,b_2)$ from the CHSH test are sent to the channel. With this strategy, one can compute a lower bound to the sum rate,
\bea
 \mathcal{L}_{s}^{(Q)} =  \max_{p(u_1)p(u_2)} I(U_1,U_2;Y). \label{eq:lb}
\eea
The right-hand-side can be computed by using the Blaut-Arimoto algorithm. Note that $ \mathcal{R}_s^{(Q)}\geq \mathcal{L}_s^{(Q)}$ for the sum rate in Eq. (\ref{eq:coml}).

Achievable quantum sum rates, serving as a lower bound to the quantum sum capacity, are computed and shown in Fig. \ref{fig:plot}. The sum rate exceeds the sum capacity with local resources for a range of the channel parameter $\eta$, thereby showing a quantum advantage.

\section{ Quantum Advantages} 
\label{sec6}

The magic square game (MSG) is defined by a tuple $G_{MSG}=(\mathcal{A}_1,\mathcal{A}_2,\mathcal{B}_1,\mathcal{B}_2,\mathcal{W}_{MS})$ \cite{Brassard2005} with
\bea
    &\mathcal{A}_1&=\mathcal{B}_1=\{0,1,2\}, \nonumber \\
    &\mathcal{A}_2&=\mathcal{B}_2=\{000,001,010,011,100,101,110,111\},\nonumber
\eea
and the winning condition $\mathcal{W}_{MS}$, which can be described as follows. Each party receives a single-bit input $a_1 \in \mathcal{A}_1$ for party $A$ and $b_1 \in \mathcal{B}_1$ for party $B$. Two parties are supposed to respond with a three-bit output 
\bea
a_2=(a^0_2,a^1_2,a^2_2)~\mathrm{and}~ b_2=(b^0_2,b^1_2,b^2_2)  \nonumber
\eea
and win the game if the following conditions are satisfied,
\begin{itemize}
\item $a^0_{2}\oplus a^1_{2}\oplus a^2_{2}=0$,
\item $b^0_{2}\oplus b^1_{2}\oplus b^2_{2}=1$, and 
\item $a^{b_{1}}_{2}=b^{a_{1}}_{2}$.
\end{itemize}
The winning probability with local resources is given by $8/9$, and quantum resources enable two parties to win the game with certainty. The magic square game is therefore a quantum pseudo-telepathy game.

From the game, $(G_{MSG},\eta_w,\eta_l)$-depolarizing game channels can be constructed, see Eq. (\ref{eq:nn}). Let $\mathcal{N}_{MSG}^{(\eta_w,\eta_l)}$ denote a MAC defined by referring to the MSG. We consider two instances of the parameterization, 
\begin{itemize}
\item $(\eta_w,\eta_l)=(1,\eta)$, and
\item $(\eta_w,\eta_l)=(\eta,1-\eta)$.
\end{itemize}
We apply Theorem \ref{thm:main} and it is straightforward to find the quantum sum capacity accordingly,
\bea
&&   \mathcal{C}^{(Q)}_{s}\left(\mathcal{N}^{(1,\eta)}_{MSG}\right) = \log 9, ~~\mathrm{and}\nonumber \\[2pt]
 &&	\mathcal{C}^{(Q)}_{s}\left(\mathcal{N}^{(\eta,1-\eta)}_{MSG}\right) = \log 9-h(\eta),\nonumber
\eea
which are also maximal over all non-signaling resources. Since local strategies do not win the MSG with certainty, it also holds that for both cases, 
\bea
\mathcal{C}^{(L)}_{s}\left(\mathcal{N}^{(\eta_w, \eta_l)}_{MSG}\right) < \mathcal{C}^{(Q)}_{s}\left(\mathcal{N}^{(\eta_w, \eta_l)}_{MSG}\right) \nonumber 
\eea
which shows the advantages of quantum resources over local ones in channel capacities. 

The cooperative encoding with quantum resources, achieving the maximal sum capacity, can be constructed from the winning strategy, which is the correlation box from the $(2,3,8)$ Bell scenario \cite{Brassard2005}. Two parties $A$ and $B$ begin by sharing four-qubit entangled states 
   \begin{align}
   	\vert\Psi\rangle_{AB}=&\frac{1}{2}\left( \vert00\rangle_{A}\vert11\rangle_{B}-\vert01\rangle_{A}\vert10\rangle_{B}\right.\nonumber \\
   	&\left.-\vert10\rangle_{A}\vert01\rangle_{B}+\vert11\rangle_{A}\vert00\rangle_{B}\right). \label{eq30}
   	\end{align}
If the question pair to the parties $(A, B)$ is $(a_{1},b_{1})$, party $A$ applies a unitary operation $\mathscr{U}_{a_{1}}$ to her two qubits and then measures in the computational basis. Party $B$ applies a unitary operation $\mathscr{V}_{b_{1}}$ to the two qubits with her and then measures in the computational basis. The unitary operations they apply are listed as follows:
\bea
   &&	\mathscr{U}_0 =\frac{1}{\sqrt{2}}
   	\begin{pmatrix*}[r]
   		i  & 0 & 0&1\\
   		0 & -i & 1&0\\
   		0 & i & 1&0\\
   		1 & 0 & 0&i\	
   	\end{pmatrix*},~~
	 	\mathscr{V}_0  =\frac{1}{2}
 	\begin{pmatrix*}[r]
 		i  & -i & 1&1\\
 		-i & -i & 1&-1\\
 		1 & 1 & -i&i\\
 		-i & i & 1&1\	
 	\end{pmatrix*}, \nonumber \\[4pt]
&&     \mathscr{U}_1 =\frac{1}{2}
   	\begin{pmatrix*}[r]
   	i & 1 & 1&i\\
   	-i & 1 & -1&i\\
   	i & 1 & -1&-i\\
   	-i & 1 & 1&-i\	
   \end{pmatrix*}, 
~~   \mathscr{V}_1 =\frac{1}{2}
 	\begin{pmatrix*}[r]
 		-1 & i & 1&i\\
 		1 & i & 1&-i\\
 		1 & -i & 1&i\\
 		-1 & -i & 1&-i\	
 	\end{pmatrix*},  \nonumber \\[4pt]
 &&  \mathscr{U}_2 =\frac{1}{2}
	\begin{pmatrix*}[r]
	-1 & -1 & -1&1\\
	1 & 1 & -1&1\\
	1 & -1 & 1&1\\
	1 & -1 & -1&-1\	
\end{pmatrix*},~  
 \mathscr{V}_2 =\frac{1}{\sqrt{2}}
 	\begin{pmatrix*}[r]
 		1 & 0 & 0&1\\
 		-1 & 0 & 0&1\\
 		0 & 1 & 1&0\\
 		0 & 1 & -1&0\	
 	\end{pmatrix*}. \nonumber \label{eq31}
\eea
  Party $A$'s two-bit outcome is assigned to $(a_2^{0},~a_2^{1})$, with $a_2^{2}$ determined to satisfy the first condition in the winning strategy. Similarly,  $B$'s two-bit outcome is assigned to $(b_2^{0},~b_2^{1})$, with $b_2^{2}$ chosen to satisfy the second condition. Under this quantum protocol, two parties win the magic square game with certainty \cite{Brassard2005}.

\section{Conclusion}
\label{sec7}

In conclusion, we have presented an information-theoretic framework for two-sender and one-receiver MACs, where senders may realize cooperative encoding by using quantum and nonlocal resources beyond shared randomness. We have shown the capacity region and a sum rate that serves as a lower bound to the sum capacity when nonclassical resources are exploited in channel coding. We have considered the construction of MACs from quantum and nonlocal games, such that MACs introduce nonclassical noise that may be compatible with quantum or nonlocal correlations beyond shared randomness. For these channels, it turns out that local resources are insufficient to achieve the maximum value in sum capacities; quantum and nonlocal resources, which achieve the winning strategies, are essential. We have shown the exact sum capacities for those MACs from games with various types of depolarization noise. Comparing the exact sum capacities and the sum rates, we have considered the CHSH and the magic square games to construct MACs accordingly and shown nonlocal and quantum advantages over local resources. 

Our results pave the way to investigating channels in network information theory, in particular, when quantum and nonlocal resources beyond shared randomness are available to senders. The general characterization of the capacity region, along with a sum rate serving as a lower bound to the sum capacity, provides a versatile theoretical tool for investigating the general usefulness of non-classical resources in network communication. Our results are not limited to a specific construction of MACs. 

From a fundamental point of view, it would be interesting to extend our results to more than two parties and investigate the relations between multi-party games, Bell inequalities, and channel capacities, where nonsignaling correlations have a much richer structure than their counterparts in two-party games and thus may resolve nonclassical noise over multiple parties. Channel capacities may also be used to characterize useful nonlocal resources for network communication.

\begin{acknowledgments} 
This work is supported by the National Research Foundation of Korea (2022M1A3C2069728, RS2024-00408613, RS-2025-00561467) and the Institute for Information \& Communication Technology Promotion (IITP) (RS-2023-00229524, RS-2025-02304540, RS-2025-25464876, RS-2025-25464616) and KAIST Quantum+X Convergence R\&D Project.

\end{acknowledgments}


\begin{thebibliography}{34}%
\makeatletter
\providecommand \@ifxundefined [1]{%
 \@ifx{#1\undefined}
}%
\providecommand \@ifnum [1]{%
 \ifnum #1\expandafter \@firstoftwo
 \else \expandafter \@secondoftwo
 \fi
}%
\providecommand \@ifx [1]{%
 \ifx #1\expandafter \@firstoftwo
 \else \expandafter \@secondoftwo
 \fi
}%
\providecommand \natexlab [1]{#1}%
\providecommand \enquote  [1]{``#1''}%
\providecommand \bibnamefont  [1]{#1}%
\providecommand \bibfnamefont [1]{#1}%
\providecommand \citenamefont [1]{#1}%
\providecommand \href@noop [0]{\@secondoftwo}%
\providecommand \href [0]{\begingroup \@sanitize@url \@href}%
\providecommand \@href[1]{\@@startlink{#1}\@@href}%
\providecommand \@@href[1]{\endgroup#1\@@endlink}%
\providecommand \@sanitize@url [0]{\catcode `\\12\catcode `\$12\catcode
  `\&12\catcode `\#12\catcode `\^12\catcode `\_12\catcode `\%12\relax}%
\providecommand \@@startlink[1]{}%
\providecommand \@@endlink[0]{}%
\providecommand \url  [0]{\begingroup\@sanitize@url \@url }%
\providecommand \@url [1]{\endgroup\@href {#1}{\urlprefix }}%
\providecommand \urlprefix  [0]{URL }%
\providecommand \Eprint [0]{\href }%
\providecommand \doibase [0]{https://doi.org/}%
\providecommand \selectlanguage [0]{\@gobble}%
\providecommand \bibinfo  [0]{\@secondoftwo}%
\providecommand \bibfield  [0]{\@secondoftwo}%
\providecommand \translation [1]{[#1]}%
\providecommand \BibitemOpen [0]{}%
\providecommand \bibitemStop [0]{}%
\providecommand \bibitemNoStop [0]{.\EOS\space}%
\providecommand \EOS [0]{\spacefactor3000\relax}%
\providecommand \BibitemShut  [1]{\csname bibitem#1\endcsname}%
\let\auto@bib@innerbib\@empty
\bibitem [{\citenamefont {Holevo}(2011)}]{Holevo2011}%
  \BibitemOpen
  \bibfield  {author} {\bibinfo {author} {\bibfnamefont {A.~S.}\ \bibnamefont
  {Holevo}},\ }\href@noop {} {\emph {\bibinfo {title} {Probabilistic and
  statistical aspects of quantum theory}}},\ Publications of the Scuola Normale
  Superiore\ (\bibinfo  {publisher} {Scuola Normale Superiore},\ \bibinfo
  {address} {Pisa, Italy},\ \bibinfo {year} {2011})\BibitemShut {NoStop}%
\bibitem [{\citenamefont {Holevo}(1998)}]{651037}%
  \BibitemOpen
  \bibfield  {author} {\bibinfo {author} {\bibfnamefont {A.}~\bibnamefont
  {Holevo}},\ }\bibfield  {title} {\bibinfo {title} {The capacity of the
  quantum channel with general signal states},\ }\href
  {https://doi.org/10.1109/18.651037} {\bibfield  {journal} {\bibinfo
  {journal} {IEEE Transactions on Information Theory}\ }\textbf {\bibinfo
  {volume} {44}},\ \bibinfo {pages} {269} (\bibinfo {year} {1998})}\BibitemShut
  {NoStop}%
\bibitem [{\citenamefont {Schumacher}\ and\ \citenamefont
  {Westmoreland}(1997)}]{PhysRevA.56.131}%
  \BibitemOpen
  \bibfield  {author} {\bibinfo {author} {\bibfnamefont {B.}~\bibnamefont
  {Schumacher}}\ and\ \bibinfo {author} {\bibfnamefont {M.~D.}\ \bibnamefont
  {Westmoreland}},\ }\bibfield  {title} {\bibinfo {title} {Sending classical
  information via noisy quantum channels},\ }\href
  {https://doi.org/10.1103/PhysRevA.56.131} {\bibfield  {journal} {\bibinfo
  {journal} {Phys. Rev. A}\ }\textbf {\bibinfo {volume} {56}},\ \bibinfo
  {pages} {131} (\bibinfo {year} {1997})}\BibitemShut {NoStop}%
\bibitem [{\citenamefont {Gisin}\ \emph {et~al.}(2002)\citenamefont {Gisin},
  \citenamefont {Ribordy}, \citenamefont {Tittel},\ and\ \citenamefont
  {Zbinden}}]{RevModPhys.74.145}%
  \BibitemOpen
  \bibfield  {author} {\bibinfo {author} {\bibfnamefont {N.}~\bibnamefont
  {Gisin}}, \bibinfo {author} {\bibfnamefont {G.}~\bibnamefont {Ribordy}},
  \bibinfo {author} {\bibfnamefont {W.}~\bibnamefont {Tittel}},\ and\ \bibinfo
  {author} {\bibfnamefont {H.}~\bibnamefont {Zbinden}},\ }\bibfield  {title}
  {\bibinfo {title} {Quantum cryptography},\ }\href
  {https://doi.org/10.1103/RevModPhys.74.145} {\bibfield  {journal} {\bibinfo
  {journal} {Rev. Mod. Phys.}\ }\textbf {\bibinfo {volume} {74}},\ \bibinfo
  {pages} {145} (\bibinfo {year} {2002})}\BibitemShut {NoStop}%
\bibitem [{\citenamefont {Bennett}\ \emph {et~al.}(1992)\citenamefont
  {Bennett}, \citenamefont {Brassard},\ and\ \citenamefont
  {Mermin}}]{PhysRevLett.68.557}%
  \BibitemOpen
  \bibfield  {author} {\bibinfo {author} {\bibfnamefont {C.~H.}\ \bibnamefont
  {Bennett}}, \bibinfo {author} {\bibfnamefont {G.}~\bibnamefont {Brassard}},\
  and\ \bibinfo {author} {\bibfnamefont {N.~D.}\ \bibnamefont {Mermin}},\
  }\bibfield  {title} {\bibinfo {title} {Quantum cryptography without bell's
  theorem},\ }\href {https://doi.org/10.1103/PhysRevLett.68.557} {\bibfield
  {journal} {\bibinfo  {journal} {Phys. Rev. Lett.}\ }\textbf {\bibinfo
  {volume} {68}},\ \bibinfo {pages} {557} (\bibinfo {year} {1992})}\BibitemShut
  {NoStop}%
\bibitem [{\citenamefont {Curty}\ \emph {et~al.}(2004)\citenamefont {Curty},
  \citenamefont {Lewenstein},\ and\ \citenamefont
  {L\"utkenhaus}}]{PhysRevLett.92.217903}%
  \BibitemOpen
  \bibfield  {author} {\bibinfo {author} {\bibfnamefont {M.}~\bibnamefont
  {Curty}}, \bibinfo {author} {\bibfnamefont {M.}~\bibnamefont {Lewenstein}},\
  and\ \bibinfo {author} {\bibfnamefont {N.}~\bibnamefont {L\"utkenhaus}},\
  }\bibfield  {title} {\bibinfo {title} {Entanglement as a precondition for
  secure quantum key distribution},\ }\href
  {https://doi.org/10.1103/PhysRevLett.92.217903} {\bibfield  {journal}
  {\bibinfo  {journal} {Phys. Rev. Lett.}\ }\textbf {\bibinfo {volume} {92}},\
  \bibinfo {pages} {217903} (\bibinfo {year} {2004})}\BibitemShut {NoStop}%
\bibitem [{\citenamefont {Ac\'{\i}n}\ and\ \citenamefont
  {Gisin}(2005)}]{PhysRevLett.94.020501}%
  \BibitemOpen
  \bibfield  {author} {\bibinfo {author} {\bibfnamefont {A.}~\bibnamefont
  {Ac\'{\i}n}}\ and\ \bibinfo {author} {\bibfnamefont {N.}~\bibnamefont
  {Gisin}},\ }\bibfield  {title} {\bibinfo {title} {Quantum correlations and
  secret bits},\ }\href {https://doi.org/10.1103/PhysRevLett.94.020501}
  {\bibfield  {journal} {\bibinfo  {journal} {Phys. Rev. Lett.}\ }\textbf
  {\bibinfo {volume} {94}},\ \bibinfo {pages} {020501} (\bibinfo {year}
  {2005})}\BibitemShut {NoStop}%
\bibitem [{\citenamefont {Ekert}(1991)}]{PhysRevLett.67.661}%
  \BibitemOpen
  \bibfield  {author} {\bibinfo {author} {\bibfnamefont {A.~K.}\ \bibnamefont
  {Ekert}},\ }\bibfield  {title} {\bibinfo {title} {Quantum cryptography based
  on bell's theorem},\ }\href {https://doi.org/10.1103/PhysRevLett.67.661}
  {\bibfield  {journal} {\bibinfo  {journal} {Phys. Rev. Lett.}\ }\textbf
  {\bibinfo {volume} {67}},\ \bibinfo {pages} {661} (\bibinfo {year}
  {1991})}\BibitemShut {NoStop}%
\bibitem [{\citenamefont {Ac\'{\i}n}\ \emph {et~al.}(2007)\citenamefont
  {Ac\'{\i}n}, \citenamefont {Brunner}, \citenamefont {Gisin}, \citenamefont
  {Massar}, \citenamefont {Pironio},\ and\ \citenamefont
  {Scarani}}]{PhysRevLett.98.230501}%
  \BibitemOpen
  \bibfield  {author} {\bibinfo {author} {\bibfnamefont {A.}~\bibnamefont
  {Ac\'{\i}n}}, \bibinfo {author} {\bibfnamefont {N.}~\bibnamefont {Brunner}},
  \bibinfo {author} {\bibfnamefont {N.}~\bibnamefont {Gisin}}, \bibinfo
  {author} {\bibfnamefont {S.}~\bibnamefont {Massar}}, \bibinfo {author}
  {\bibfnamefont {S.}~\bibnamefont {Pironio}},\ and\ \bibinfo {author}
  {\bibfnamefont {V.}~\bibnamefont {Scarani}},\ }\bibfield  {title} {\bibinfo
  {title} {Device-independent security of quantum cryptography against
  collective attacks},\ }\href {https://doi.org/10.1103/PhysRevLett.98.230501}
  {\bibfield  {journal} {\bibinfo  {journal} {Phys. Rev. Lett.}\ }\textbf
  {\bibinfo {volume} {98}},\ \bibinfo {pages} {230501} (\bibinfo {year}
  {2007})}\BibitemShut {NoStop}%
\bibitem [{\citenamefont {Bennett}\ \emph {et~al.}(1999)\citenamefont
  {Bennett}, \citenamefont {Shor}, \citenamefont {Smolin},\ and\ \citenamefont
  {Thapliyal}}]{PhysRevLett.83.3081}%
  \BibitemOpen
  \bibfield  {author} {\bibinfo {author} {\bibfnamefont {C.~H.}\ \bibnamefont
  {Bennett}}, \bibinfo {author} {\bibfnamefont {P.~W.}\ \bibnamefont {Shor}},
  \bibinfo {author} {\bibfnamefont {J.~A.}\ \bibnamefont {Smolin}},\ and\
  \bibinfo {author} {\bibfnamefont {A.~V.}\ \bibnamefont {Thapliyal}},\
  }\bibfield  {title} {\bibinfo {title} {Entanglement-assisted classical
  capacity of noisy quantum channels},\ }\href
  {https://doi.org/10.1103/PhysRevLett.83.3081} {\bibfield  {journal} {\bibinfo
   {journal} {Phys. Rev. Lett.}\ }\textbf {\bibinfo {volume} {83}},\ \bibinfo
  {pages} {3081} (\bibinfo {year} {1999})}\BibitemShut {NoStop}%
\bibitem [{\citenamefont {Nayak}(1999)}]{10.5555/795665.796491}%
  \BibitemOpen
  \bibfield  {author} {\bibinfo {author} {\bibfnamefont {A.}~\bibnamefont
  {Nayak}},\ }\bibfield  {title} {\bibinfo {title} {Optimal lower bounds for
  quantum automata and random access codes},\ }in\ \href@noop {} {\emph
  {\bibinfo {booktitle} {Proceedings of the 40th Annual Symposium on
  Foundations of Computer Science}}},\ \bibinfo {series and number} {FOCS '99}\
  (\bibinfo  {publisher} {IEEE Computer Society},\ \bibinfo {address} {USA},\
  \bibinfo {year} {1999})\ p.\ \bibinfo {pages} {369}\BibitemShut {NoStop}%
\bibitem [{\citenamefont {Brunner}\ \emph {et~al.}(2014)\citenamefont
  {Brunner}, \citenamefont {Cavalcanti}, \citenamefont {Pironio}, \citenamefont
  {Scarani},\ and\ \citenamefont {Wehner}}]{Brunner2014}%
  \BibitemOpen
  \bibfield  {author} {\bibinfo {author} {\bibfnamefont {N.}~\bibnamefont
  {Brunner}}, \bibinfo {author} {\bibfnamefont {D.}~\bibnamefont {Cavalcanti}},
  \bibinfo {author} {\bibfnamefont {S.}~\bibnamefont {Pironio}}, \bibinfo
  {author} {\bibfnamefont {V.}~\bibnamefont {Scarani}},\ and\ \bibinfo {author}
  {\bibfnamefont {S.}~\bibnamefont {Wehner}},\ }\bibfield  {title} {\bibinfo
  {title} {Bell nonlocality},\ }\href
  {https://doi.org/10.1103/RevModPhys.86.419} {\bibfield  {journal} {\bibinfo
  {journal} {Rev. Mod. Phys.}\ }\textbf {\bibinfo {volume} {86}},\ \bibinfo
  {pages} {419} (\bibinfo {year} {2014})}\BibitemShut {NoStop}%
\bibitem [{\citenamefont {Buhrman}\ \emph
  {et~al.}(2010{\natexlab{a}})\citenamefont {Buhrman}, \citenamefont {Cleve},
  \citenamefont {Massar},\ and\ \citenamefont {de~Wolf}}]{Buhrman2010}%
  \BibitemOpen
  \bibfield  {author} {\bibinfo {author} {\bibfnamefont {H.}~\bibnamefont
  {Buhrman}}, \bibinfo {author} {\bibfnamefont {R.}~\bibnamefont {Cleve}},
  \bibinfo {author} {\bibfnamefont {S.}~\bibnamefont {Massar}},\ and\ \bibinfo
  {author} {\bibfnamefont {R.}~\bibnamefont {de~Wolf}},\ }\bibfield  {title}
  {\bibinfo {title} {Nonlocality and communication complexity},\ }\href
  {https://doi.org/10.1103/revmodphys.82.665} {\bibfield  {journal} {\bibinfo
  {journal} {Reviews of Modern Physics}\ }\textbf {\bibinfo {volume} {82}},\
  \bibinfo {pages} {665} (\bibinfo {year} {2010}{\natexlab{a}})}\BibitemShut
  {NoStop}%
\bibitem [{\citenamefont {Quek}\ and\ \citenamefont
  {Shor}(2017)}]{QueckShor2017}%
  \BibitemOpen
  \bibfield  {author} {\bibinfo {author} {\bibfnamefont {Y.}~\bibnamefont
  {Quek}}\ and\ \bibinfo {author} {\bibfnamefont {P.~W.}\ \bibnamefont
  {Shor}},\ }\bibfield  {title} {\bibinfo {title} {Quantum and superquantum
  enhancements to two-sender, two-receiver channels},\ }\href
  {https://doi.org/10.1103/PhysRevA.95.052329} {\bibfield  {journal} {\bibinfo
  {journal} {Phys. Rev. A}\ }\textbf {\bibinfo {volume} {95}},\ \bibinfo
  {pages} {052329} (\bibinfo {year} {2017})}\BibitemShut {NoStop}%
\bibitem [{\citenamefont {Yun}\ \emph {et~al.}(2020)\citenamefont {Yun},
  \citenamefont {Rai},\ and\ \citenamefont {Bae}}]{YunRaiBae2020}%
  \BibitemOpen
  \bibfield  {author} {\bibinfo {author} {\bibfnamefont {J.}~\bibnamefont
  {Yun}}, \bibinfo {author} {\bibfnamefont {A.}~\bibnamefont {Rai}},\ and\
  \bibinfo {author} {\bibfnamefont {J.}~\bibnamefont {Bae}},\ }\bibfield
  {title} {\bibinfo {title} {Nonlocal network coding in interference
  channels},\ }\href {https://doi.org/10.1103/PhysRevLett.125.150502}
  {\bibfield  {journal} {\bibinfo  {journal} {Phys. Rev. Lett.}\ }\textbf
  {\bibinfo {volume} {125}},\ \bibinfo {pages} {150502} (\bibinfo {year}
  {2020})}\BibitemShut {NoStop}%
\bibitem [{\citenamefont {Leditzky}\ \emph {et~al.}(2020)\citenamefont
  {Leditzky}, \citenamefont {Alhejji}, \citenamefont {Levin},\ and\
  \citenamefont {Smith}}]{Leditzky2020}%
  \BibitemOpen
  \bibfield  {author} {\bibinfo {author} {\bibfnamefont {F.}~\bibnamefont
  {Leditzky}}, \bibinfo {author} {\bibfnamefont {M.~A.}\ \bibnamefont
  {Alhejji}}, \bibinfo {author} {\bibfnamefont {J.}~\bibnamefont {Levin}},\
  and\ \bibinfo {author} {\bibfnamefont {G.}~\bibnamefont {Smith}},\ }\bibfield
   {title} {\bibinfo {title} {Playing games with multiple access channels},\
  }\href {https://doi.org/10.1038/s41467-020-15240-w} {\bibfield  {journal}
  {\bibinfo  {journal} {Nature Communications}\ }\textbf {\bibinfo {volume}
  {11}},\ \bibinfo {pages} {1497} (\bibinfo {year} {2020})}\BibitemShut
  {NoStop}%
\bibitem [{\citenamefont {Seshadri}\ \emph {et~al.}(2023)\citenamefont
  {Seshadri}, \citenamefont {Leditzky}, \citenamefont {Siddhu},\ and\
  \citenamefont {Smith}}]{Seshadri2023}%
  \BibitemOpen
  \bibfield  {author} {\bibinfo {author} {\bibfnamefont {A.}~\bibnamefont
  {Seshadri}}, \bibinfo {author} {\bibfnamefont {F.}~\bibnamefont {Leditzky}},
  \bibinfo {author} {\bibfnamefont {V.}~\bibnamefont {Siddhu}},\ and\ \bibinfo
  {author} {\bibfnamefont {G.}~\bibnamefont {Smith}},\ }\bibfield  {title}
  {\bibinfo {title} {On the separation of correlation-assisted sum capacities
  of multiple access channels},\ }\href
  {https://doi.org/10.1109/TIT.2023.3274434} {\bibfield  {journal} {\bibinfo
  {journal} {IEEE Transactions on Information Theory}\ }\textbf {\bibinfo
  {volume} {69}},\ \bibinfo {pages} {5805} (\bibinfo {year}
  {2023})}\BibitemShut {NoStop}%
\bibitem [{\citenamefont {Rezaeian}\ and\ \citenamefont
  {Grant}(2004)}]{Rezaeian_Grant_AB_algoMAC_2004}%
  \BibitemOpen
  \bibfield  {author} {\bibinfo {author} {\bibfnamefont {M.}~\bibnamefont
  {Rezaeian}}\ and\ \bibinfo {author} {\bibfnamefont {A.}~\bibnamefont
  {Grant}},\ }\bibfield  {title} {\bibinfo {title} {Computation of total
  capacity for discrete memoryless multiple-access channels},\ }\href
  {https://doi.org/10.1109/TIT.2004.836661} {\bibfield  {journal} {\bibinfo
  {journal} {IEEE Transactions on Information Theory}\ }\textbf {\bibinfo
  {volume} {50}},\ \bibinfo {pages} {2779} (\bibinfo {year}
  {2004})}\BibitemShut {NoStop}%
\bibitem [{\citenamefont {Watanabe}\ and\ \citenamefont
  {Kamoi}(2009)}]{Watanabe_Kamoi_2002_AB_algoMAC_Article}%
  \BibitemOpen
  \bibfield  {author} {\bibinfo {author} {\bibfnamefont {Y.}~\bibnamefont
  {Watanabe}}\ and\ \bibinfo {author} {\bibfnamefont {K.}~\bibnamefont
  {Kamoi}},\ }\bibfield  {title} {\bibinfo {title} {A formulation of the
  channel capacity of multiple-access channel},\ }\href
  {https://doi.org/10.1109/TIT.2009.2015991} {\bibfield  {journal} {\bibinfo
  {journal} {IEEE Transactions on Information Theory}\ }\textbf {\bibinfo
  {volume} {55}},\ \bibinfo {pages} {2083} (\bibinfo {year}
  {2009})}\BibitemShut {NoStop}%
\bibitem [{\citenamefont {Ahlswede}(1973)}]{Ahlswede}%
  \BibitemOpen
  \bibfield  {author} {\bibinfo {author} {\bibfnamefont {R.}~\bibnamefont
  {Ahlswede}},\ }\bibfield  {title} {\bibinfo {title} {Multi-way communication
  channels}\ }(\bibinfo {year} {1973})\BibitemShut {NoStop}%
\bibitem [{\citenamefont {Liao}(1972)}]{Liao}%
  \BibitemOpen
  \bibfield  {author} {\bibinfo {author} {\bibfnamefont {H.}~\bibnamefont
  {Liao}},\ }\bibfield  {title} {\bibinfo {title} {Multiple access channels},\
  }\href@noop {} {\bibfield  {journal} {\bibinfo  {journal} {Phd thesis,
  Department of Electrical Engineering, University of Hawaii}\ } (\bibinfo
  {year} {1972})}\BibitemShut {NoStop}%
\bibitem [{\citenamefont {Yeung}(2002)}]{yeungFirstCourseInformation2002}%
  \BibitemOpen
  \bibfield  {author} {\bibinfo {author} {\bibfnamefont {R.~W.}\ \bibnamefont
  {Yeung}},\ }\href {https://doi.org/10.1007/978-1-4419-8608-5} {\emph
  {\bibinfo {title} {A {{First Course}} in {{Information Theory}}}}},\ edited
  by\ \bibinfo {editor} {\bibfnamefont {J.~K.}\ \bibnamefont {Wolf}},\
  Information {{Technology}}: {{Transmission}}, {{Processing}} and {{Storage}}\
  (\bibinfo  {publisher} {Springer US},\ \bibinfo {address} {Boston, MA},\
  \bibinfo {year} {2002})\BibitemShut {NoStop}%
\bibitem [{\citenamefont {B{\"u}hler}\ and\ \citenamefont
  {Wunder}(2011)}]{Bhler2011ANO}%
  \BibitemOpen
  \bibfield  {author} {\bibinfo {author} {\bibfnamefont {J.}~\bibnamefont
  {B{\"u}hler}}\ and\ \bibinfo {author} {\bibfnamefont {G.}~\bibnamefont
  {Wunder}},\ }\bibfield  {title} {\bibinfo {title} {A note on capacity
  computation for the discrete multiple access channel},\ }\href
  {https://api.semanticscholar.org/CorpusID:2263877} {\bibfield  {journal}
  {\bibinfo  {journal} {IEEE Transactions on Information Theory}\ }\textbf
  {\bibinfo {volume} {57}},\ \bibinfo {pages} {1906} (\bibinfo {year}
  {2011})}\BibitemShut {NoStop}%
\bibitem [{\citenamefont {Pereg}\ \emph {et~al.}(2025)\citenamefont {Pereg},
  \citenamefont {Deppe},\ and\ \citenamefont {Boche}}]{pereg2025}%
  \BibitemOpen
  \bibfield  {author} {\bibinfo {author} {\bibfnamefont {U.}~\bibnamefont
  {Pereg}}, \bibinfo {author} {\bibfnamefont {C.}~\bibnamefont {Deppe}},\ and\
  \bibinfo {author} {\bibfnamefont {H.}~\bibnamefont {Boche}},\ }\bibfield
  {title} {\bibinfo {title} {The {{Multiple-Access Channel With Entangled
  Transmitters}}},\ }\href {https://doi.org/10.1109/TIT.2024.3516507}
  {\bibfield  {journal} {\bibinfo  {journal} {IEEE Transactions on Information
  Theory}\ }\textbf {\bibinfo {volume} {71}},\ \bibinfo {pages} {1096}
  (\bibinfo {year} {2025})}\BibitemShut {NoStop}%
\bibitem [{\citenamefont {Gamal}\ and\ \citenamefont
  {Kim}(2012)}]{Gamal_and_Kim}%
  \BibitemOpen
  \bibfield  {author} {\bibinfo {author} {\bibfnamefont {A.~E.}\ \bibnamefont
  {Gamal}}\ and\ \bibinfo {author} {\bibfnamefont {Y.-H.}\ \bibnamefont
  {Kim}},\ }\href@noop {} {\emph {\bibinfo {title} {Network Information
  Theory}}}\ (\bibinfo  {publisher} {Cambridge University Press},\ \bibinfo
  {address} {New York},\ \bibinfo {year} {2012})\BibitemShut {NoStop}%
\bibitem [{\citenamefont {Greenberger}\ \emph {et~al.}(1989)\citenamefont
  {Greenberger}, \citenamefont {Horne},\ and\ \citenamefont {Zeilinger}}]{GHZ}%
  \BibitemOpen
  \bibfield  {author} {\bibinfo {author} {\bibfnamefont {D.~M.}\ \bibnamefont
  {Greenberger}}, \bibinfo {author} {\bibfnamefont {M.~A.}\ \bibnamefont
  {Horne}},\ and\ \bibinfo {author} {\bibfnamefont {A.}~\bibnamefont
  {Zeilinger}},\ }\bibinfo {title} {Going beyond bell's theorem},\ in\ \href
  {https://doi.org/10.1007/978-94-017-0849-4_10} {\emph {\bibinfo {booktitle}
  {Bell's Theorem, Quantum Theory and Conceptions of the Universe}}},\ \bibinfo
  {editor} {edited by\ \bibinfo {editor} {\bibfnamefont {M.}~\bibnamefont
  {Kafatos}}}\ (\bibinfo  {publisher} {Springer Netherlands},\ \bibinfo
  {address} {Dordrecht},\ \bibinfo {year} {1989})\ pp.\ \bibinfo {pages}
  {69--72}\BibitemShut {NoStop}%
\bibitem [{\citenamefont {Mermin}(1990)}]{Mermin}%
  \BibitemOpen
  \bibfield  {author} {\bibinfo {author} {\bibfnamefont {N.~D.}\ \bibnamefont
  {Mermin}},\ }\bibfield  {title} {\bibinfo {title} {Quantum mysteries
  revisited},\ }\href@noop {} {\bibfield  {journal} {\bibinfo  {journal}
  {American Journal of Physics}\ }\textbf {\bibinfo {volume} {58}},\ \bibinfo
  {pages} {731} (\bibinfo {year} {1990})}\BibitemShut {NoStop}%
\bibitem [{\citenamefont {Clauser}\ \emph {et~al.}(1969)\citenamefont
  {Clauser}, \citenamefont {Horne}, \citenamefont {Shimony},\ and\
  \citenamefont {Holt}}]{chsh}%
  \BibitemOpen
  \bibfield  {author} {\bibinfo {author} {\bibfnamefont {J.~F.}\ \bibnamefont
  {Clauser}}, \bibinfo {author} {\bibfnamefont {M.~A.}\ \bibnamefont {Horne}},
  \bibinfo {author} {\bibfnamefont {A.}~\bibnamefont {Shimony}},\ and\ \bibinfo
  {author} {\bibfnamefont {R.~A.}\ \bibnamefont {Holt}},\ }\bibfield  {title}
  {\bibinfo {title} {Proposed experiment to test local hidden-variable
  theories},\ }\href {https://doi.org/10.1103/PhysRevLett.23.880} {\bibfield
  {journal} {\bibinfo  {journal} {Phys. Rev. Lett.}\ }\textbf {\bibinfo
  {volume} {23}},\ \bibinfo {pages} {880} (\bibinfo {year} {1969})}\BibitemShut
  {NoStop}%
\bibitem [{\citenamefont {Eisert}\ \emph {et~al.}(1999)\citenamefont {Eisert},
  \citenamefont {Wilkens},\ and\ \citenamefont {Lewenstein}}]{EWL1999}%
  \BibitemOpen
  \bibfield  {author} {\bibinfo {author} {\bibfnamefont {J.}~\bibnamefont
  {Eisert}}, \bibinfo {author} {\bibfnamefont {M.}~\bibnamefont {Wilkens}},\
  and\ \bibinfo {author} {\bibfnamefont {M.}~\bibnamefont {Lewenstein}},\
  }\bibfield  {title} {\bibinfo {title} {Quantum games and quantum
  strategies},\ }\href {https://doi.org/10.1103/PhysRevLett.83.3077} {\bibfield
   {journal} {\bibinfo  {journal} {Phys. Rev. Lett.}\ }\textbf {\bibinfo
  {volume} {83}},\ \bibinfo {pages} {3077} (\bibinfo {year}
  {1999})}\BibitemShut {NoStop}%
\bibitem [{\citenamefont {Brandenburger}\ and\ \citenamefont
  {La~Mura}(2016)}]{Brandenburger_2016}%
  \BibitemOpen
  \bibfield  {author} {\bibinfo {author} {\bibfnamefont {A.}~\bibnamefont
  {Brandenburger}}\ and\ \bibinfo {author} {\bibfnamefont {P.}~\bibnamefont
  {La~Mura}},\ }\bibfield  {title} {\bibinfo {title} {Team decision problems
  with classical and quantum signals},\ }\href
  {https://doi.org/10.1098/rsta.2015.0096} {\bibfield  {journal} {\bibinfo
  {journal} {Philosophical Transactions of the Royal Society A: Mathematical,
  Physical and Engineering Sciences}\ }\textbf {\bibinfo {volume} {374}},\
  \bibinfo {pages} {20150096} (\bibinfo {year} {2016})}\BibitemShut {NoStop}%
\bibitem [{\citenamefont {Koniorczyk}\ \emph {et~al.}(2020)\citenamefont
  {Koniorczyk}, \citenamefont {Bodor},\ and\ \citenamefont
  {Pint\'er}}]{Koniorczyk+2020}%
  \BibitemOpen
  \bibfield  {author} {\bibinfo {author} {\bibfnamefont {M.}~\bibnamefont
  {Koniorczyk}}, \bibinfo {author} {\bibfnamefont {A.}~\bibnamefont {Bodor}},\
  and\ \bibinfo {author} {\bibfnamefont {M.}~\bibnamefont {Pint\'er}},\
  }\bibfield  {title} {\bibinfo {title} {Ex ante versus ex post equilibria in
  classical bayesian games with a nonlocal resource},\ }\href
  {https://doi.org/10.1103/PhysRevA.101.062115} {\bibfield  {journal} {\bibinfo
   {journal} {Phys. Rev. A}\ }\textbf {\bibinfo {volume} {101}},\ \bibinfo
  {pages} {062115} (\bibinfo {year} {2020})}\BibitemShut {NoStop}%
\bibitem [{\citenamefont {Auletta}\ \emph {et~al.}(2021)\citenamefont
  {Auletta}, \citenamefont {Ferraioli}, \citenamefont {Rai}, \citenamefont
  {Scarpa},\ and\ \citenamefont {Winter}}]{Auletta+2021}%
  \BibitemOpen
  \bibfield  {author} {\bibinfo {author} {\bibfnamefont {V.}~\bibnamefont
  {Auletta}}, \bibinfo {author} {\bibfnamefont {D.}~\bibnamefont {Ferraioli}},
  \bibinfo {author} {\bibfnamefont {A.}~\bibnamefont {Rai}}, \bibinfo {author}
  {\bibfnamefont {G.}~\bibnamefont {Scarpa}},\ and\ \bibinfo {author}
  {\bibfnamefont {A.}~\bibnamefont {Winter}},\ }\bibfield  {title} {\bibinfo
  {title} {Belief-invariant and quantum equilibria in games of incomplete
  information},\ }\href
  {https://doi.org/https://doi.org/10.1016/j.tcs.2021.09.041} {\bibfield
  {journal} {\bibinfo  {journal} {Theoretical Computer Science}\ }\textbf
  {\bibinfo {volume} {895}},\ \bibinfo {pages} {151} (\bibinfo {year}
  {2021})}\BibitemShut {NoStop}%
\bibitem [{\citenamefont {Brassard}\ \emph {et~al.}(2005)\citenamefont
  {Brassard}, \citenamefont {Broadbent},\ and\ \citenamefont
  {Tapp}}]{Brassard2005}%
  \BibitemOpen
  \bibfield  {author} {\bibinfo {author} {\bibfnamefont {G.}~\bibnamefont
  {Brassard}}, \bibinfo {author} {\bibfnamefont {A.}~\bibnamefont
  {Broadbent}},\ and\ \bibinfo {author} {\bibfnamefont {A.}~\bibnamefont
  {Tapp}},\ }\bibfield  {title} {\bibinfo {title} {Quantum pseudo-telepathy},\
  }\href {https://doi.org/10.1007/s10701-005-7353-4} {\bibfield  {journal}
  {\bibinfo  {journal} {Foundations of Physics}\ }\textbf {\bibinfo {volume}
  {35}},\ \bibinfo {pages} {1877} (\bibinfo {year} {2005})}\BibitemShut
  {NoStop}%
\end{thebibliography}

%

\end{document}